\documentclass[12pt]{iopart}
\usepackage{iopams}
\expandafter\let\csname equation*\endcsname\relax
\expandafter\let\csname endequation*\endcsname\relax
\usepackage{amsmath}
\usepackage[]{graphicx} 

\begin{document}
\title[On the equivalence of self-consistent equations]{On the equivalence of self-consistent equations for nonuniform liquids: a unified description of the various modifications}

\author {Hiroshi Frusawa\footnote[1]{e-mail: frusawa.hiroshi@kochi-tech.ac.jp}}
\address{Laboratory of Statistical Physics, Kochi University of Technology, Tosa-Yamada, Kochi 782-8502, Japan.}

\begin{abstract}
A variety of self-consistent (SC) equations have been proposed for non-uniform states of liquid particles under external fields, including adsorbed states at solid substrates and confined states in pores.
External fields represent not only confining geometries but also fixed solutes.
We consider SC equations ranging from the modified Poisson-Boltzmann equations for the Coulomb potential to the hydrostatic linear response equation for the equilibrium density distribution of Lennard-Jones fluids.
Here, we present a unified equation that explains the apparent diversity of previous forms and proves the equivalence of various SC equations.
This unified description of SC equations is obtained from a hybrid method combining the conventional density functional theory and statistical field theory.
The Gaussian approximation of density fluctuations around a mean-field distribution is performed based on the developed hybrid framework, allowing us to derive a novel form of the grand-potential density functional that provides the unified SC equation for equilibrium density.
\end{abstract}

\section{Introduction}
Many studies on dense uniform liquids have demonstrated the dominant role of short-ranged harshly repulsive intermolecular forces in determining structural correlations, which provides the insight that  attractive intermolecular forces essentially cancel due to the symmetric configurations of molecules in uniform liquids [1-3].
In contrast, translational invariance of molecular arrangements is broken in non-uniform liquids [4-12].
Accordingly, both attractive and repulsive forces can, for instance, significantly influence the structure of a Lennard-Jones (LJ) fluid next to a hard wall because the vector sums of the long-range attractive forces do not cancel each other.

There are several situations in which such non-uniform states of liquid molecules emerge.
Inhomogeneities occur [4-12]: (i) at interfaces between liquid-gas, liquid-liquid, and liquid-crystal phases; (ii) in the adsorption of liquids at solid substrates or walls; (iii) for fluids in confining geometries, such as slits and pores; and (iv) in the sedimentation equilibrium of colloidal particles under gravity.
Theoretically, a non-uniform state can be created by adding an interaction potential term associated with an external field.
For example, an external field may represent the presence of walls, slits, or pores [4-14].
We can also find adequate fields that are not only responsible for shifts and rounding of bulk phase transitions, but also induce wetting or drying transitions and related phenomena such as capillary condensation.
Furthermore, an external field is able to describe the interactions of fixed solutes with solvent liquids that significantly perturb the solvent density around solutes [4-14].

Based on this theoretical approach, we can evaluate the solvation free energy per solute and effective interaction potential between solutes [1, 4-14].
Specifically, the solvation thermodynamics in charged and polar systems underlies important processes in biology, including protein folding and ligand binding \cite{chandler nature,weeks solvation}.
This indicates that the Coulomb potential as well as the LJ potential is significant when considering the long-range part of external field.
In fact, among the theoretical issues related to external Coulomb fields are the formulations of the following diverse phenomena:
(i) hydration thermodynamics of hydrophobic and ionic solutes \cite{chandler nature}; (ii) solvation of highly charged colloids \cite{weeks solvation}; (iii) unconventional phenomena such as Coulomb attraction between like-charged surfaces and overcharging of solvated objects due to the accumulation of counterions in solvents (see \cite{naji} for a review); and (iv) novel properties observed around the interfaces between electrodes and room temperature ionic liquids, including the crossover from overscreening to crowding with an increase in applied voltage and the long-range underscreening detected by surface force measurements [16-19].

All of these phenomena associated with Coulomb interactions stem from the strong correlations between charged particles, which requires going beyond the Poisson-Boltzmann (PB) equation, a typical mean-field equation for calculating the Coulomb potential induced by charge-charge interactions \cite{naji}.
A variety of modified PB equations [14-34], proposed so far, represent a subset of the self-consistent (SC) equations considering intermolecular correlations.
Our concern here is with the following two types of the modified PB equations.
One type is derived from the charge smearing model [14, 20-24, 35-40] and the other adds higher-order gradient terms to the original PB equation [15-18, 25-30].
Both types of modifications have coexisted as complementary to each other, and the direct connections of these different types of the modified PB equations with the liquid-state theory, or the vast knowledge on correlation functions [1, 41], remain to be clarified.

Meanwhile, for dense non-Coulombic liquids under the influence of external fields, there is another set of SC equations for predicting the non-uniform distribution of average density $\rho_{\mathrm{eq}}({\bf r})$ in equilibrium [4, 5, 14, 41-47].
This type of SC theory can be traced back to the van der Waals interface theory \cite{weeks review,lcw chandler}.
The essential ingredient of the classical van der Waals theory is that gas-liquid coexistence can be attributed to intermolecular competition between short-range repulsions and long-range attractions, thereby providing a qualitatively accurate description of slowly varying interfaces.
A recent SC equation modified the van der Waals equation to quantitatively describe local interfacial structures on the molecular scale \cite{weeks review}.

The goal of this paper is to show that there is an underlying framework to yield various SC equations, despite their apparently diverse forms.
To this end, we demonstrate the equivalence of the aforementioned SC equations for both the equilibrium density and Coulomb potential, using a hybrid method \cite{frusawa2,frusawa3} that combines the density functional theory (DFT) [6-12] and statistical field theory [15, 23, 31-34, 36, 50-52].

The DFT incorporates as input microscopic correlations due to the inter-particle interactions [6-12], thus providing a fruitful approach to the descriptions of either microscopic structures in inhomogeneous fluids or states of various matters, including not only hard spheres, but also soft-core particles, mixtures and anisotropic particles.
The statistical field theory [15, 23, 31-34, 36, 50-52], on the other hand, can incorporate fluctuations, especially important in the vicinity of the critical point, into the formulation in systematic ways using the Gaussian approximation, many-loop expansion, variation method, and the renormalization group theory.
Benefiting from both of the DFT and statistical field theory, the hybrid theory \cite{frusawa2,frusawa3} allows us to incorporate knowledge from the liquid-state theory [1, 41] into a field-theoretic treatment of density fluctuations in inhomogeneous fluids.
Accordingly, a unified equation for $\rho_{\mathrm{eq}}({\bf r})$ can be obtained from a novel form of the grand-potential density functional.
The unified equation corresponds to an extension from the equation based on the Percus' test particle method [47, 53] for uniform liquids to that for non-uniform liquids.
It will also be demonstrated that the single unified equation for $\rho_{\mathrm{eq}}({\bf r})$ can yield a variety of existing SC equations [4, 5, 14-34, 41-47].

In the remainder of this paper, we first present a review of previous SC equations [4, 5, 14-34, 41-47] and our main results related to the unification discussed above:
sections 2 and 3 provide the formal background and essential results for demonstrating the equivalence between previous SC equations.
In Sections 4 and 5, comparisons are made between previous SC equations and the obtained forms derived from a single unified equation for $\rho_{\mathrm{eq}}({\bf r})$:
we investigate the SC equations for non-Coulombic liquids [4, 5, 14, 41-47] in Section 4, whereas Section 5 presents a unified equation for $\rho_{\mathrm{eq}}({\bf r})$ expressed by the Coulomb potential for verifying the equivalence of modified PB equations [14-34].
After demonstrating the relevance of our hybrid framework \cite{frusawa2,frusawa3} for SC equations, Section 6 outlines the derivation scheme for the grand-potential density functional, a key functional to obtain the unified SC equation.
Summary and conclusions are given in Section 7.
In the appendices, full details of the formulations regarding not only the present results but also the previous modified PB equations can be found.

\section{Formal background}
\subsection{Target systems in terms of the potential separation [1-5, 14, 20, 21, 46, 47]}
In non-uniform liquids, there are two types of interaction potentials: one is the interaction potential between liquid particles, which produces intermolecular forces in both uniform and non-uniform liquids, whereas the other is a potential that creates a known external field felt by constituent particles, irrespective of liquid states, due to the presence of fixed solutes, confining geometries, and so on.
The former interaction potential $v({\bf r})$, as a function of the intermolecular distance $|{\bf r}|\equiv r$, includes the hard-sphere (HS) potential, LJ potential, and Coulomb potential.
The latter external potential $J({\bf r})$, which is uniquely determined by the location ${\bf r}$ of particles, is caused not only by the existence of a hard wall, but also by long-range repulsive or attractive particle-wall interactions, such as Coulomb and LJ interactions.
It should be noted that not only $v({\bf r})$ and $J({\bf r})$ but also the other energetic quantities, appearing in this paper, are defined in the $k_BT$-unit.

Let us introduce the liquid systems through explaining the potential splitting into two parts:
\begin{flalign}
v({\bf r})&=v_0({\bf r})+v_1({\bf r}),
\label{interaction split}\\
J({\bf r})&=J_0({\bf r})+J_1({\bf r}).
\label{external split}
\end{flalign}
In general, $v_0$ and $J_0$ denote harshly repulsive parts and $v_1$ and $J_1$ slowly varying parts.
In a dense and uniform fluid, the subscript "0" indicates that the potentials are those in a reference system that has been found sufficient for determining the essential structural properties [1-3].
As the first approximation in treating $J_0({\bf r})$, we have estimated from $J_0$ an effective radius $R$ of fixed object, such as a fixed solute, that excludes constituent liquid particles (see eq. (\ref{appendix radius}) for the definition of the effective radius $R$) \cite{weeks review,lcw chandler}.

We investigate the SC equations for HS fluids, attractive particle systems, and point-charge systems as listed in Table 1.
Correspondingly, we give concrete forms of the interaction potential splitting ($v=v_0+v_1$) for the HS fluids, LJ ones and point-charge systems.
The so-called WCA separation, proposed by Weeks, Chandler and Andersen, applies to attractive particle systems including the LJ fluids [1-3]. 

First, in the HS fluids, $v_0$ is identified with the full potential $v_{\mathrm{HS}}$ of HS interactions (i.e., $v_0=v_{\mathrm{HS}}$), and it follows that $v_1\equiv 0$.
Next, the WCA separation for the LJ fluids creates the WCA potential $v_{\mathrm{WCA}}({\bf r})$ arising from the LJ potential $v_{LJ}$ truncated at the minimum as follows:
\begin{flalign}
v_0({\bf r})=v_{\mathrm{WCA}}({\bf r})\equiv
\left\{
\begin{array}{l}
v_{LJ}({\bf r})+\epsilon_{LJ}\quad\mathrm{if}\>r\leq r_m\\
\\
0\quad\qquad\qquad\>\>\,\mathrm{if}\>r>r_m,
\end{array}
\right.
\label{v0 LJ}
\end{flalign}
and
\begin{flalign}
v_1({\bf r})=
\left\{
\begin{array}{l}
-\epsilon_{LJ}\quad\qquad\>\>\mathrm{if}\>r\leq r_m\\
\\
v_{LJ}({\bf r})\quad\qquad\mathrm{if}\>r>r_m.
\end{array}
\right.
\label{v1 LJ}
\end{flalign}
In eqs. (\ref{v0 LJ}) and (\ref{v1 LJ}), $r_m$ and $\epsilon_{LJ}$ denote the separation and depth of potential minimum, respectively.
Equation (\ref{v0 LJ}) is a typical form of the WCA potential in attractive particle systems.

Lastly, the point-charge systems have the Coulomb interaction potential that is always repulsive:
\begin{equation}
v({\bf r})=v_{\mathrm{el}}({\bf r})\equiv \frac{z^2l_B}{r},
\label{Coulomb interaction}
\end{equation}
where we consider particles carrying charges of $+ze$ and the Bjerrum length $l_B$ is defined as the distance at which two unit charges have interaction energy $k_BT$.
Despite the long-range nature, there is no characteristic length for the potential splitting; nevertheless, previous studies have proposed the potential splitting [14, 20, 21] such that
\begin{flalign}
v_{\mathrm{el}}({\bf r})&=v_0({\bf r})+v_1({\bf r}),
\label{Coulomb separation}\\
v_0({\bf r})&=z^2l_B\frac{\mathrm{erfc}(r/\xi)}{r},
\label{Coulomb v0}\\
v_1({\bf r})&=z^2l_B\frac{\mathrm{erf}(r/\xi)}{r},
\label{Coulomb v1}
\end{flalign}
which has been demonstrated to be useful in finding the non-uniform Coulomb potential by adjusting the characteristic length $\xi$.
General forms of $v_1({\bf r})$ in point-charge systems are given in Appendix A.2 where it is clarified that this type of potential splitting in the Coulomb potential is equivalent to the introduction of charge smearing model (or Onsager charge smearing) for strongly-coupled Coulomb systems [14, 20-24, 35-40].

\begin{table}
\caption{A list of previous self-consistent equations derived from a unified equation (\ref{tcf equilibrium density}) [EL$\,=\,$Euler-Lagrange; RY$\,=\,$Ramakrishnan-Yussouff; HLR$\,=\,$hydrostatic linear response; PB$\,=\,$Poisson-Boltzmann].\\}
\footnotesize
\begin{tabular}{c|ccc}
Constituents&
\begin{tabular}{l}
Self-consistently\\
determined variables
\end{tabular}&
Original equation&
Modified equation\\[3pt]
\hline\hline\\
Hard spheres&Equilibrium density&
\begin{tabular}{l}
EL equation of\\
the RY functional
\end{tabular}
&\begin{tabular}{c}
HLR equation\\
(eq. (\ref{weeks hlr}))
\end{tabular}\\[18pt]
Attractive particles&
\begin{tabular}{c}
Equilibrium density\\[4pt]
Reference density
\end{tabular}&
\begin{tabular}{c}
{\itshape ditto}\\[4pt]
---
\end{tabular}
&\begin{tabular}{c}
{\itshape ditto}\\[4pt]
Mean-field equation
\end{tabular}\\
&&&(eq. (\ref{mean-field reference}))\\[3pt]
\hline\\
Point charges&
Coulomb potential&
PB equation&
\begin{tabular}{l}
Finite-spread PB equation
\end{tabular}\\
&&&(eqs. (\ref{weeks PB}), (\ref{appendix poisson1}) and (\ref{appendix poisson2}))\\[10pt]
Point charges&
Coulomb potential&
PB equation&
\begin{tabular}{c}
Higher-order PB equation
\end{tabular}\\
&&&(eqs. (\ref{higher-order pb-lue}) and (\ref{appendix another higher-order pb}))
\end{tabular}
\end{table}
\subsection{A variety of previous self-consistent equations}
We outline previous SC equations [4, 5, 14-34, 42-47] in the above three systems, which this paper aims to derive in a unified manner.
Table 1 lists the previous SC equations for these systems. 
In Table 1, the SC equations are classified into two groups:
the former group can determine the equilibrium density for the HS and LJ fluids [4, 5, 14, 42-47], whereas the latter group modifies the PB equation to improve the mean-field solution of Coulomb potential [14-34].

{\itshape Conventional mean-field equation for the reference density $\rho_{\mathrm{ref}}({\bf r})$ [4, 5, 14, 42-47].}---
The conventional mean-field equation is used when considering a slowly varying interaction potential $v_1({\bf r})$ in a non-uniform liquid with the external potential $J({\bf r})$ applied.
The solution of the mean-field equation has been referred to as the reference density $\rho_{\mathrm{ref}}({\bf r})$.
In the outside of the excluded region with the effective radius of $R$ \cite{weeks review,lcw chandler}, which is defined below in eq. (\ref{appendix radius}), $\rho_{\mathrm{ref}}({\bf r})$ satisfies
\begin{flalign}
\rho_{\mathrm{ref}}({\bf r})
=e^{\mu-J_1({\bf r})-\int d{\bf r}'v_{1}({\bf r}-{\bf r}')\rho_{\mathrm{ref}}({\bf r}')},
\label{mean-field reference}
\end{flalign}
where $\mu$ denotes the chemical potential in the $k_BT$-unit.
It is found from eq. (\ref{mean-field reference}) that the spatial dependence of $\rho_{\mathrm{ref}}({\bf r})$ for HS fluids arises only from $J_1({\bf r})$ because of $v_1({\bf r})\equiv 0$.

The usual form relating $J_0({\bf r})$ to $R$ is
\begin{equation}
R=\int_0^{\infty}dr\,\left\{
1-e^{-J_0({\bf r})}
\right\},
\label{appendix radius}
\end{equation}
which is a measure of the region $w$ that excludes the liquid particles due to the harshly repulsive field $J_0({\bf r})$.
Considering that the reference density $\rho_{\mathrm{ref}}({\bf r})$ necessarily vanishes inside the region $w$ due to the presence of $J_0({\bf r})$, the reference density is expressed by
\begin{flalign}
\rho_{\mathrm{ref}}({\bf r})=
\left\{
\begin{array}{l}
0;\qquad{\bf r}\in w,\\
\\
\rho_{\mathrm{ref}}^0\,e^{-\Delta \phi({\bf r})};\>{\bf r}\notin w,
\end{array}
\right.
\label{appendix reference solute}
\end{flalign}
where the reference density $\rho_{\mathrm{ref}}^0\equiv\rho_{\mathrm{ref}}({\bf r}_0)$ at a specified position of ${\bf r}_0$ is treated separately, and the shifted chemical potential $\mu_{\mathrm{ref}}$, defined by
\begin{flalign}
\rho_{\mathrm{ref}}^0&\equiv e^{\mu_{\mathrm{ref}}}=e^{\mu-\phi({\bf r}_0)},
\label{shifted chemical potential}\\
\phi({\bf r}_0)&=J_1({\bf r}_0)+\int d{\bf r}'v_1({\bf r}_0-{\bf r}')\rho_{\mathrm{ref}}({\bf r}'),
\label{appendix att ext}
\end{flalign}
is utilized for obtaining the equilibrium density, according to the hydrostatic approximation [4, 5, 43-45].
Accordingly, the reference density at other locations is given by eq. (\ref{appendix reference solute}) using $\rho_{\mathrm{ref}}^0$ and the external potential difference $\Delta\phi({\bf r})\equiv \phi({\bf r})-\phi({\bf r}_0)$ from the external field given by eq. (\ref{appendix att ext}).

Previous studies based on the hydrostatic approximation have demonstrated that the mean-field like treatment works well for the purpose of obtaining the equilibrium density distribution of a dense non-uniform liquid with the help of the above reference density $\rho_{\mathrm{ref}}^0$. 

{\itshape The SC equation for the equilibrium density $\rho_{\mathrm{eq}}({\bf r})$ [4, 5, 42-47].}---
In attractive particle systems as well as in HS fluids, the equilibrium density $\rho_{\mathrm{eq}}({\bf r})$ is obtained from evaluating the deviation from the reference density $\rho_{\mathrm{ref}}^0$ at a specified position of ${\bf r}_0$.
The evaluation is based on the approximation of the hydrostatic linear response (HLR) [4, 5, 14, 42-47], which reads for the equilibrium density at ${\bf r}_0$
\begin{eqnarray}
\rho_{\mathrm{eq}}({\bf r}_0)=\rho_{\mathrm{ref}}^0\left[
1+\int d{\bf r}'\,c_{\mathrm{0}}({\bf r}_0-{\bf r}')\left\{
\rho_{\mathrm{eq}}({\bf r}')-\rho_{\mathrm{ref}}^0
\right\}\right],
\label{weeks hlr}
\end{eqnarray}
using the direct correlation function (DCF) $c_0({\bf r})$ for a uniform system of liquid particles interacting via the harshly repulsive potential $v_0({\bf r})$.
The DCF depends on the density; in eq. (\ref{weeks hlr}), however, we use the DCF at a uniform density of $\rho^0_{\mathrm{ref}}$.
The same equation (\ref{weeks hlr}) applies to both of attractive particle systems and HS fluids because the HLR equation [4, 5, 14, 42-47] is formulated for $v_0({\bf r})$, irrespective of the systems considered.
Equation (\ref{weeks hlr}) is reduced to the closure in the hypernetted chain (HNC) approximation [1] when the density difference $\rho_{\mathrm{eq}}({\bf r})-\rho_{\mathrm{ref}}^0$ is related to the total correlation function (TCF) $h_0({\bf r})$ in the $v_0$--interacting systems: $\rho_{\mathrm{eq}}({\bf r})-\rho_{\mathrm{ref}}^0=\rho_{\mathrm{ref}}^0h_0({\bf r}-{\bf r}_0)$, according to the Percus' test particle method [47, 53].
In eq. (\ref{weeks hlr}), we take into account the non-uniformity by incorporating the contribution of slowly varying potential $v_1({\bf r})$ into the mean-field solution $\rho_{\mathrm{ref}}^0$ as given by eq. (\ref{mean-field reference}).

{\itshape Modified PB equations for the Coulomb potential $\Psi({\bf r})$ [14-30].}---
Lastly, we mention two types of modified PB equations (see Appendix A for the details).
Both modifications are relevant to describe the charge-charge correlations between charged particles in the one component plasma (OCP) at strong coupling, concentrated electrolytes, and room temperature ionic liquids.
It has also been found that these two modifications explain the deviation from the PB solution for counterion distribution over a wide range of from the intermediate to strong coupling [14-30].

One type is the finite-spread PB equation [14, 20-24] based on the charge smearing model.
In the uniform OCP, a smeared charge distribution is simply characterized by the mean distance $a$ between adjacent charges and has been found useful to represent the strong repulsive correlations between the charges;
for non-uniform systems, $a$ is supposed to represent the mean separation in the ground state (see Appendix A.1 for the details of the definition).
In the local molecular field theory [14, 20, 21], the Gaussian charge smearing model provides
\begin{flalign}
\nabla^2\Psi ({\bf r})
&=-4\pi l_B\overline{\overline{\,e^{\mu-J_1({\bf r})-\Psi ({\bf r})}}}
\label{weeks PB}\\
\overline{\overline{\,\mathcal{O}({\bf r})}}
&=\frac{1}{\{\pi(a/m )^2\}^{3/2}}\int d{\bf r}'
\exp\left\{-\left(\frac{m |{\bf r}-{\bf r}'|}{a}\right)^2\right\}
\mathcal{O}({\bf r}'),
\label{weeks gaussian}
\end{flalign}
where a constant $m$ is an adjusting parameter (see also Appendix A.3);
for instance, $m=1/1.4$ for the uniform OCP and $m=1/0.6$ for the counterion system in the presence of one charged wall, according to previous results at strong coupling [14, 20, 21].

We would also like to investigate the other type of higher-order PB equations [15-18, 25-30].
This type of modification adds higher-order gradient terms to the PB equation such that \cite{lue}
\begin{flalign}
\nabla^2\left\{1-\frac{a^2}{4}\nabla^2+\frac{a^4}{16}\nabla^4\right\}\Psi({{\bf r}})
=-4\pi l_B z^2e^{\mu-J_1({\bf r})-\Psi ({\bf r})};
\label{higher-order pb-lue}
\end{flalign}
see Appendix A.4 for the details of higher-order PB equations [15-18, 25-30].
The other type of higher-order PB equation, which is truncated at the biharmonic $\nabla^2\nabla^2$--term, has been referred to as the Bazant-Storey-Kornyshev (BSK) equation [16-18] in binary systems such as concentrated electrolytes and room temperature ionic liquids.
\section{Central results for proving the equivalence of previous self-consistent equations}
\subsection{Grand-potential density functional $\widetilde{\Omega}[\rho_{\mathrm{ref}}]$}
As detailed in Section 6, the density-functional integral representation of the grand potential is a powerful tool to evaluate fluctuating density fields around the reference density $\rho_{\mathrm{ref}}$ with the help of the conventional DFT [6-12].
The hybrid method \cite{frusawa2,frusawa3} combining the DFT [6-12] and statistical field theory [15, 23, 31-34, 36, 50-52] allows us to obtain the grand-potential density functional $\widetilde{\Omega}[\rho_{\mathrm{ref}}]$ into which the additional contribution due to density fluctuations is incorporated.
The resulting form is
\begin{flalign}
\widetilde{\Omega}[\rho_{\mathrm{ref}}]&=\widetilde{\Omega_0}[\rho_{\mathrm{ref}}]+U_1[\rho_{\mathrm{ref}}],
\label{omega sum}\\
\widetilde{\Omega_0}[\rho_{\mathrm{ref}}]
&=\mathcal{F}_0^{\mathrm{ex}}[\rho^0_{\mathrm{ref}}]-\frac{1}{2}\iint d{\bf r}d{\bf r}'
\Delta\rho_{\mathrm{ref}}({\bf r})h_0({\bf r}-{\bf r}')\Delta\rho_{\mathrm{ref}}({\bf r}')
\nonumber\\
&\qquad\qquad+\int d{\bf r}\left[
\rho_{\mathrm{ref}}({\bf r})\ln\rho_{\mathrm{ref}}({\bf r})-\rho_{\mathrm{ref}}({\bf r})
-\rho_{\mathrm{ref}}({\bf r})\mu
\right],
\label{omega0 result}\\
U_1[\rho_{\mathrm{ref}}]&=\frac{1}{2}\iint d{\bf r}d{\bf r}'
\rho_{\mathrm{ref}}({\bf r})v_{\mathrm{att}}({\bf r}-{\bf r}')\rho_{\mathrm{ref}}({\bf r}')
+\int d{\bf r}\,\rho_{\mathrm{ref}}({\bf r})J_1({\bf r}).
\label{v1 interaction energy}
\end{flalign}
In eq. (\ref{omega0 result}), $\mathcal{F}_0^{\mathrm{ex}}[\rho^0_{\mathrm{ref}}]$ represents the excess part of the {\itshape intrinsic} Helmholtz free energy, $\Delta\rho_{\mathrm{ref}}({\bf r})=\rho_{\mathrm{ref}}({\bf r})-\rho_{\mathrm{ref}}^0$ and the reference TCF $h_0({\bf r})$ is related to the reference DCF $c_0({\bf r})$ through the Ornstein-Zernike equation given below.
The subscript "0" in the TCF have double meanings as well as that in the DCF: the TCF is not only the function at a uniform density, which is equal to a reference density $\rho_{\mathrm{ref}}^0=\rho_{\mathrm{ref}}({\bf r}_0)$ at ${\bf r}_0$, but is also the function in repulsive $v_0$--interaction systems.
The subscript "1" on the right hand side (rhs) of eq. (\ref{v1 interaction energy}) is altered to the subscript "att", clarifying that {\itshape the potential splitting given by eq. (\ref{interaction split}) applies only to the attractive particle systems in our results}: $v=v_0+v_{\mathrm{att}}$.

Equation (\ref{omega0 result}) indicates that the repulsive part of the grand potential $\widetilde{\Omega_0}$ is composed of two parts:
the interaction energy term and the ideal-gas contribution.
The second term on the rhs of eq. (\ref{omega0 result}), representing the interaction energy, is expressed using the TCF instead of the DCF, other than the conventional DFT [6-12].
In eq. (\ref{v1 interaction energy}), on the other hand, the interaction energy due to the slowly varying attractive part $v_{\mathrm{att}}({\bf r})$ of interaction potential is evaluated in the mean-field approximation.
Such treatment is apparently a crude estimation but has been found to be relevant for evaluating the long-range contribution to $\widetilde{\Omega}$  in non-uniform liquids [14, 46, 47];
actually, it is quite difficult to find density-density correlations by explicitly considering the non-uniformity.

\subsection{Reference and equilibrium densities}
As seen from the expression (\ref{v1 interaction energy}), we regard only the contribution due to the long-range attractive interactions as the additional part of the grand-potential density functional $\widetilde{\Omega}[\rho_{\mathrm{ref}}]$.
Correspondingly, in this study, $v_0({\bf r})$ reads for the following three systems:
\begin{flalign}
v_0({\bf r})=
\left\{
\begin{array}{l}
v_{\mathrm{WCA}}({\bf r})\quad(\mathrm{Attractive\,particles})\\
\\
v_{\mathrm{HS}}({\bf r})\quad\quad(\mathrm{Hard\,spheres})\\
\\
v_{\mathrm{el}}({\bf r})\quad\quad\>(\mathrm{Point\,charges}).
\end{array}
\right.
\label{our v0}
\end{flalign}
In what follows, the same rule of the subscript "0" as above applies to the subscript "0" in the TCF ($h_0({\bf r})$) and DCF ($c_0({\bf r})$).
Equation (\ref{mean-field reference}) with the above expression (\ref{our v0}) implies that the spatial dependence of $\rho_{\mathrm{ref}}({\bf r})$ is determined solely by the external potential $J_1({\bf r})$ for both of the HS fluids and point-charge systems because of $v_{\mathrm{att}}({\bf r})\equiv 0$.

Once the external potential $J_1$ is treated as a variable field $\mathcal{J}$, we can differentiate the grand potential with respect to $\mathcal{J}$, thereby generating the equilibrium density $\rho_{\mathrm{eq}}({\bf r})$ with the help of a variable reference density $\rho_{\mathrm{ref}}^{\mathcal{J}}({\bf r})\equiv e^{\mu-\mathcal{J}({\bf r})-\int d{\bf r}'v_{\mathrm{att}}({\bf r}-{\bf r}')\rho_{\mathrm{ref}}({\bf r}')}$ as follows:
\begin{flalign}
\rho_{\mathrm{eq}}({\bf r})&=
\left.
\frac{\delta\widetilde{\Omega}[\rho^{\mathcal{J}}_{\mathrm{ref}}]}{\delta \mathcal{J}({\bf r})}
\right|_{\mathcal{J}=J_1}\nonumber\\
&=\left.
\frac{\delta\rho^{\mathcal{J}}_{\mathrm{ref}}({\bf r})}{\delta\mathcal{J}({\bf r})}\right|_{\mathcal{J}=J_1}
\left.
\frac{\delta\widetilde{\Omega}[\rho^{\mathcal{J}}_{\mathrm{ref}}]}{\delta \rho^{\mathcal{J}}_{\mathrm{ref}}({\bf r})}
\right|_{\rho^{\mathcal{J}}_{\mathrm{ref}}=\rho_{\mathrm{ref}}}\nonumber\\
&=-\rho_{\mathrm{ref}}({\bf r})\left.
\frac{\delta\widetilde{\Omega}[\rho^{\mathcal{J}}_{\mathrm{ref}}]}{\delta \rho^{\mathcal{J}}_{\mathrm{ref}}({\bf r})}\right|_{\rho^{\mathcal{J}}_{\mathrm{ref}}=\rho_{\mathrm{ref}}},
\label{general equilibrium density}
\end{flalign}
where use has been made of the following relation in the last line of the above equation:
\begin{flalign}
\left.
\frac{\delta\rho^{\mathcal{J}}_{\mathrm{ref}}({\bf r})}{\delta\mathcal{J}({\bf r})}\right|_{\mathcal{J}=J_1}
&=
\left.
\frac{\delta}{\delta\mathcal{J}}\left\{
 e^{\mu-\mathcal{J}({\bf r})-\int d{\bf r}'v_{\mathrm{att}}({\bf r}-{\bf r}')\rho_{\mathrm{ref}}({\bf r}')}
\right\}
\right|_{\mathcal{J}=J_1}\nonumber\\
&=-e^{\mu-J_1({\bf r})-\int d{\bf r}'v_{\mathrm{att}}({\bf r}-{\bf r}')\rho_{\mathrm{ref}}({\bf r}')}.
\label{differentiating reference density}
\end{flalign}
It also follows from eqs. (\ref{omega sum}) to (\ref{v1 interaction energy}) that the functional differentiation of $\widetilde{\Omega}[\rho_{\mathrm{ref}}]$ with respect to $\rho_{\mathrm{ref}}({\bf r})$ yields
\begin{flalign}
\left.
\frac{\delta\widetilde{\Omega}[\rho^{\mathcal{J}}_{\mathrm{ref}}]}{\delta \rho^{\mathcal{J}}_{\mathrm{ref}}({\bf r})}\right|_{\rho^{\mathcal{J}}_{\mathrm{ref}}=\rho_{\mathrm{ref}}}
&=-\int d{\bf r}'\,h_0({\bf r}-{\bf r}')\Delta\rho_{\mathrm{ref}}({\bf r}')\nonumber\\
&\qquad\underline{
+\int d{\bf r}'\,v_{\mathrm{att}}({\bf r}-{\bf r}')\rho_{\mathrm{ref}}({\bf r}')
+J_1({\bf r})-\mu+\ln\rho_{\mathrm{ref}}({\bf r})}
+\rho_{\mathrm{ref}}({\bf r})\left.
\frac{\delta\mathcal{J}}{\delta\rho^{\mathcal{J}}_{\mathrm{ref}}}
\right|_{\rho^{\mathcal{J}}_{\mathrm{ref}}=\rho_{\mathrm{ref}}}
\nonumber\\
&=-\int d{\bf r}'\,h_0({\bf r}-{\bf r}')\Delta\rho_{\mathrm{ref}}({\bf r}')-1,
\label{appendix differentiation}
\end{flalign}
where the resulting form seen in the last line has been obtained from both using eq. (\ref{differentiating reference density}) and canceling the underlined terms due to the relation (\ref{mean-field reference}).
Combination of eqs. (\ref{general equilibrium density}) and (\ref{appendix differentiation}) provides the single unified equation for $\rho_{\mathrm{eq}}({\bf r})$:
\begin{flalign}
\rho_{\mathrm{eq}}({\bf r})=\rho_{\mathrm{ref}}({\bf r})\left\{
1+\int d{\bf r}'\,h_0({\bf r}-{\bf r}')\Delta\rho_{\mathrm{ref}}({\bf r}')
\right\},
\label{tcf equilibrium density}
\end{flalign}
which is our main result.
Plugging the Ornstein-Zernike equation into eq. (\ref{tcf equilibrium density}), the HLR equation (\ref{weeks hlr}) [4, 5, 14, 42-47] will be obtained in the next section.

\section{Comparison with the HLR equation (\ref{weeks hlr}) for the equilibrium density $\rho_{\mathrm{eq}}({\bf r})$}
We use the Ornstein-Zernike equation \cite{hansen},
\begin{flalign}
h_0({\bf r}-{\bf r}')=c_0({\bf r}-{\bf r}')+\int d{\bf r}"\,
c_0({\bf r}-{\bf r}")\rho_{\mathrm{ref}}^0\,h_0({\bf r}"-{\bf r}'),
\label{inhomo oz}
\end{flalign}
for the reference density $\rho_{\mathrm{ref}}^0$ given by eq. (\ref{shifted chemical potential}) because both of the DCF and TCF, $c_0({\bf r})$ and $h_0({\bf r})$, are defined for uniform liquids with the density of $\rho_{\mathrm{ref}}^0$.
Plugging eq. (\ref{inhomo oz}) into eq. (\ref{tcf equilibrium density}) at ${\bf r}_0$, we have
\begin{flalign}
&\frac{\rho_{\mathrm{eq}}({\bf r}_0)-\rho_{\mathrm{ref}}^0}{\rho_{\mathrm{ref}}^0}=
\int d{\bf r}'\,h_0({\bf r}_0-{\bf r}')\Delta\rho_{\mathrm{ref}}({\bf r}')\nonumber\\
&=\int d{\bf r}'\,c_0({\bf r}_0-{\bf r}')\Delta\rho_{\mathrm{ref}}({\bf r}')
+\int d{\bf r}"\,c_0({\bf r}_0-{\bf r}")
\rho_{\mathrm{ref}}^0\int d{\bf r}'
h_0({\bf r}"-{\bf r}')\Delta\rho_{\mathrm{ref}}({\bf r}').
\label{pre oz equilibrium}
\end{flalign}
The main result (\ref{tcf equilibrium density}) is again inserted into $\rho_{\mathrm{ref}}^0\int d{\bf r}'
h_0({\bf r}"-{\bf r}')\Delta\rho_{\mathrm{ref}}({\bf r}')$ of the above expression with the relation $\rho_{\mathrm{ref}}({\bf r}")=\rho_{\mathrm{ref}}^0+\Delta\rho_{\mathrm{ref}}({\bf r}")$, yielding
\begin{flalign}
\rho_{\mathrm{ref}}^0\int d{\bf r}'
h_0({\bf r}"-{\bf r}')\Delta\rho_{\mathrm{ref}}({\bf r}')
&=\rho_{\mathrm{ref}}^0\left\{
\frac{\rho_{\mathrm{eq}}({\bf r}")-\rho_{\mathrm{ref}}({\bf r}")}{\rho_{\mathrm{ref}}^0+\Delta\rho_{\mathrm{ref}}({\bf r}")}
\right\}\nonumber\\
&\approx\rho_{\mathrm{eq}}({\bf r}")-\rho_{\mathrm{ref}}({\bf r}")
-\left\{\rho_{\mathrm{eq}}({\bf r}")-\rho_{\mathrm{ref}}({\bf r}")\right\}
\frac{\Delta\rho_{\mathrm{ref}}({\bf r}")}{\rho_{\mathrm{ref}}^0}\nonumber\\
&\approx\rho_{\mathrm{eq}}({\bf r}")-\rho_{\mathrm{ref}}({\bf r}").
\label{underline density}
\end{flalign}
It will be seen in Section 6 and Appendix B.2 that the above approximation used in eq. (\ref{underline density}) is consistent with the approximation in deriving the grand-potential density functional $\widetilde{\Omega}[\rho_{\mathrm{ref}}]$ given by eqs. (\ref{omega sum}) to (\ref{v1 interaction energy}). 
Thus, we find that $\widetilde{\Omega}[\rho_{\mathrm{ref}}]$ leads to the HLR equation (\ref{weeks hlr}) via eq. (\ref{tcf equilibrium density}) [4, 5, 14, 42-47]:
it follows from eq. (\ref{underline density}) that eq. (\ref{pre oz equilibrium}) reduces to
\begin{flalign}
\frac{\rho_{\mathrm{eq}}({\bf r}_0)-\rho_{\mathrm{ref}}^0}{\rho_{\mathrm{ref}}^0}
&\approx\int d{\bf r}'\,c_0({\bf r}_0-{\bf r}')\Delta\rho_{\mathrm{ref}}({\bf r}')
+\int d{\bf r}"\,c_0({\bf r}_0-{\bf r}")\left\{
{\rho_{\mathrm{eq}}({\bf r}")-\rho_{\mathrm{ref}}({\bf r}")}
\right\}\nonumber\\
&=\int d{\bf r}'\,c_0({\bf r}_0-{\bf r}')\left\{
\rho_{\mathrm{eq}}({\bf r}')-\rho_{\mathrm{ref}}^0
\right\},
\label{oz equilibrium}
\end{flalign}
which is equal to the HLR equation (\ref{weeks hlr}).

\section{Comparison with modified PB equations for the Coulomb potential $\Psi ({\bf r})$}
Before comparing the unified equation (\ref{tcf equilibrium density}) for $\rho_{\mathrm{eq}}({\bf r})$ with the modified PB equations such as eqs. (\ref{weeks PB}) to (\ref{higher-order pb-lue}), we would like to see that the essential equation (\ref{tcf equilibrium density}) implies the Boltzmann distribution of $\rho_{\mathrm{eq}}({\bf r})$.
Actually, in the approximation of $1+x\approx e^x$, eq. (\ref{tcf equilibrium density}) reads for point-charge systems
\begin{flalign}
\rho_{\mathrm{eq}}({\bf r})
&\approx\rho_{\mathrm{ref}}({\bf r})e^{-\Psi ({\bf r})},
\label{equilibrium el density}\\
\Psi ({\bf r})&=-\int d{\bf r}'\,h_{\mathrm{el}}({\bf r}-{\bf r}')\Delta\rho_{\mathrm{ref}}({\bf r}'),
\label{equilibrium el potential}
\end{flalign}
where $h_{\mathrm{el}}({\bf r})$ denotes the electrostatic TCF of point charges interacting via $v_{\mathrm{el}}({\bf r})$ without the potential splitting.
This section investigates the modified PB equations, or the SC equations for the Coulomb potential $\Psi ({\bf r})$ created by Coulomb interactions between point charges.
We consider two types of the modified PB equations: the finite-spread PB equations [14, 20-24] based on the charge smearing model and the higher-order PB equations  [15-18, 25-30].
We show below that both forms can be derived from the Poisson equation for $\Psi ({\bf r})$ defined by eq. (\ref{equilibrium el potential}) using the electrostatic TCF $h_{\mathrm{el}}({\bf r})$.

\subsection{Relation between the Coulomb potential $\Psi({\bf r})$ and the DCF $c_{\mathrm{el}}({\bf r})$}
Plugging the Ornstein-Zernike equation (\ref{inhomo oz}) into eq. (\ref{equilibrium el potential}) as before, we have
\begin{flalign}
\Psi ({\bf r})
&=-\int d{\bf r}'\,c_{\mathrm{el}}({\bf r}-{\bf r}')\Delta\rho_{\mathrm{ref}}({\bf r}')
-\iint d{\bf r}'d{\bf r}"\,c_{\mathrm{el}}({\bf r}-{\bf r}")\rho_{\mathrm{ref}}({\bf r}")
h_{\mathrm{el}}({\bf r}"-{\bf r}')\Delta\rho_{\mathrm{ref}}({\bf r}')\nonumber\\
&=-\int d{\bf r}'c_{\mathrm{el}}({\bf r}-{\bf r}')\rho_{\mathrm{ref}}({\bf r}')
-\int d{\bf r}"\,c_{\mathrm{el}}({\bf r}-{\bf r}")\rho_{\mathrm{ref}}({\bf r}")\Psi ({\bf r}")-\Psi_c.
\label{oz potential}
\end{flalign}
In the last line of eq. (\ref{oz potential}), the constant potential $\Psi_c$ is related to the isothermal compressibility $\kappa_T$ as $\Psi_c=\left(k_BT\rho_{\mathrm{ref}}^0\kappa_T\right)^{-1}-1$ because of the following relation:
\begin{flalign}
1-\rho_{\mathrm{ref}}^0\int d{\bf r}\,c_{\mathrm{el}}({\bf r})
=\frac{1}{k_BT\rho_{\mathrm{ref}}^0\kappa_T}.
\label{compressibility}
\end{flalign}
The resulting expressions of the Poisson equation for the above potential $\Psi$ vary, according to the treatments of the Laplace equation for the electrostatic DCF $c_{\mathrm{el}}({\bf r})$.

To see the different expressions of the Laplace equation, it is convenient to introduce the Fourier-transform of $c_{\mathrm{el}}({\bf r})$.
In general, $c_{\mathrm{el}}({\bf k})$ can be written as
\begin{equation}
c_{\mathrm{el}}({\bf k})=-z^2l_B\frac{f ({\bf k})}{k^2},
\label{dcf el}
\end{equation}
where $f ({\bf k})$ take various forms depending on the approximations adopted:
\begin{flalign}
f ({\bf k})=
\left\{
\begin{array}{l}
e^{-\frac{(ka)^2}{4m }}\quad\qquad(\mathrm{HNC}\> \cite{ng})\\
\\
\left\{
\frac{3}{ka}j_1(ka)
\right\}^2\quad(\mathrm{Soft\,MSA}\> [22, 39, 40])\\
\\ 
\frac{1}{1+(ka)^2/4}\quad\qquad(\mathrm{Bessel\,smoothed\,model}\> [35]),
\end{array}
\right.
\label{form factor}
\end{flalign}
where $k=|{\bf k}|$ and $a$ is the mean separation between adjacent charges as mentioned before eq. (\ref{weeks PB}) (see also Appendix A.1 for the detailed definition of $a$).
In eq. (\ref{form factor}), we use $m =1.08$ according to the HNC approximation \cite{ng}, the soft MSA [22, 39, 40] signifies the mean spherical approximation (MSA) [1, 41] for soft spheres, and the smeared interaction potential is identified with the DCF in the Bessel smoothed model [35] that supposes a smeared charge distribution expressed by the modified Bessel function $K_1$ (see Appendix A.2 for the details).

\subsection{Finite-spread PB equation [14, 20-24}
In the real-space representation of the electrostatic DCF given by eq. (\ref{dcf el}), we have
\begin{flalign}
-\nabla^2\int d{\bf r}'c_{\mathrm{el}}({\bf r}-{\bf r}')\rho_{\mathrm{ref}}({\bf r}')
&=\iint d{\bf s}d{\bf r}'\,\nabla^2 v_{\mathrm{el}}({\bf r}-{\bf s})f ({\bf s}-{\bf r}')\rho_{\mathrm{ref}}({\bf r}')\nonumber\\
&=-4\pi z^2l_B\int d{\bf s}\,\delta({\bf r}-{\bf s})
\left<\rho_{\mathrm{ref}}({\bf s})\right>\nonumber\\
&=-4\pi z^2l_B\left<\rho_{\mathrm{ref}}({\bf r})\right>,
\label{dcf poisson}
\end{flalign}
where we have introduced the smeared density, $\left<\rho_{\mathrm{ref}}({\bf r})\right>$, defined by
\begin{flalign}
\left<\rho_{\mathrm{ref}}({\bf s})\right>
&=\int d{\bf r}'\,
f ({\bf s}-{\bf r}')\,\rho_{\mathrm{ref}}({\bf r}')
\nonumber\\
&=\int d{\bf r}'\,
f ({\bf s}-{\bf r}')\,e^{\mu-J_1({\bf r}')}.
\label{alpha weighted density}
\end{flalign}
From eqs. (\ref{oz potential}) and (\ref{dcf poisson}), we find
\begin{flalign}
\nabla^2\Psi({\bf r})
&=-4\pi z^2l_B\left<\rho_{\mathrm{ref}}({\bf r})\right>
+4\pi z^2l_B\left<\rho_{\mathrm{ref}}({\bf r})\right>\Psi({\bf r})
\label{linear pb}\\
&\approx-4\pi z^2l_B\left<\rho_{\mathrm{eq}}({\bf r})\right>,
\label{smeared mpb}
\end{flalign}
where the approximation (\ref{equilibrium el density}) has been applied to the above second line:
\begin{flalign}
\left<\rho_{\mathrm{eq}}({\bf r})\right>
&=\left<\rho_{\mathrm{ref}}({\bf r})e^{-\Psi({\bf r})}\right>\nonumber\\
&=\int d{\bf s}\,
f ({\bf r}-{\bf s})\,\rho_{\mathrm{ref}}({\bf s})e^{-\Psi({\bf s})};
\label{smeared equilibrium density}
\end{flalign}
see also Appendix A.3 and A.5 for the details.
It is found from eqs. (\ref{dcf poisson}) and (\ref{alpha weighted density}) that the introduction of smeared charge model is compatible with the use of the DCF in the liquid-state theory, as has been verified for the strongly coupled OCP (see Appendix A.5 for the details) \cite{frydel2016,rosenfeld gelbart,rosenfeld evans}.
Equations (\ref{alpha weighted density}), (\ref{smeared mpb}) and (\ref{smeared equilibrium density}) are reduced to eqs. (\ref{weeks PB}) and (\ref{weeks gaussian}) in the HNC approximation that provides $f ({\bf r}-{\bf s})=\{\pi(a/m )^2\}^{-3/2}\,e^{-\left(m |{\bf r}-{\bf s}|/a\right)^2}$;
however, a discrepancy exists between the values of $m$: $m=1.08$ for the HNC \cite{ng}, whereas the local molecular field theory [14, 20, 21] set that $m=1/0.6\approx 1.67$ for a counterion system as mentioned after eq. (\ref{weeks gaussian}).
Considering that the effective mean distance between point charges is diminished to $a/m$ by the factor $1/m$ in the Gaussian charge smearing model, it is suggested from the fitting result ($m\approx 1.67$) of the local molecular theory that the effective mean distance $a/m$ should be shorter than that of the HNC approximation of the uniform OCP model for a quantitative description of inhomogeneous counterion distribution in perpendicular direction to oppositely charged surface.

\subsection{Higher-order Poisson-Boltzmann equations  [15-18, 25-30]}
The higher-order PB equations add higher-order gradient terms to the Laplacian appearing in the usual Poisson equation (see also Appendix A.4 for the details).
The Fourier-space representation of eqs. (\ref{smeared mpb}) and (\ref{smeared equilibrium density}) is
\begin{flalign}
-k^2\Psi(k)=-4\pi l_B z^2f ({\bf k})\rho_{\mathrm{ref}}(-{\bf k})e^{-\Psi({\bf k})}.
\label{k mpb}
\end{flalign}
While $f ({\bf k})\rho(-{\bf k})$ can be interpreted as a smeared density when leaving $f ({\bf k})$ on the right hand side, the higher-order PB equations are generated by investigating the contribution of $f ^{-1}({\bf k})$ on the left hand side.
In other words, this subsection focuses on how to handle $k^2/f ({\bf k})$.

In the long-wave approximation valid for $ka\ll 1$, we have
\begin{flalign}
\frac{1}{f ({\bf k})}\approx
\left\{
\begin{array}{l}
1+\frac{(ka)^2}{4m}\quad\qquad(\mathrm{HNC})\\
\\
1+\frac{(ka)^2}{5}\quad\quad(\mathrm{Soft\,MSA})\\
\\ 
1+\frac{(ka)^2}{4}\quad\qquad(\mathrm{Bessel}),
\end{array}
\right.
\label{inverse f}
\end{flalign}
In the above results, $m=1.08$ as before, whereas the above approximate form for the soft MSA \cite{frydel2016,rosenfeld gelbart,rosenfeld evans} is derived as follows: we have used the approximation that
\begin{equation}
\left\{
\frac{3}{ka}j_1(ka)
\right\}^{-2}
\approx
1+\frac{(ka)^2}{5},
\label{approximate homogeneous}
\end{equation}
expanding the Bessel function $j_1(x)$ as
\begin{flalign}
j_1(x)&=\frac{\sin x-x\cos x}{x^2}\nonumber\\
&\approx \frac{x}{3}\left(
1-\frac{x^2}{10}
\right),
\label{appendix bessel}
\end{flalign}
where use has been made of the approximation,
\begin{flalign}
\sin x-x\cos x
\approx \frac{x^3}{3}-\frac{x^5}{30},
\label{appendix sin cos}
\end{flalign}
based on the expansions such that $\sin x\approx x-x^3/6+x^5/120$ and $\cos x\approx 1-x^2/2+x^4/24$.

It follows from eq. (\ref{inverse f}) that
\begin{flalign}
-\frac{k^2}{f({\bf k})}\approx
-\left\{
1+(\gamma a)^2k^2
\right\}k^2,
\label{left had side of mpb}
\end{flalign}
with various values of $\gamma$:
\begin{flalign}
\gamma=
\left\{
\begin{array}{l}
\frac{1}{2\sqrt{m}}\quad(\mathrm{HNC})\\
\\
\frac{1}{\sqrt{5}}\quad(\mathrm{Soft\,MSA})\\
\\ 
\frac{1}{2}\quad(\mathrm{Bessel}).
\end{array}
\right.
\label{gamma}
\end{flalign}
It is found from eqs. (\ref{k mpb}), (\ref{left had side of mpb}) and (\ref{gamma}) that eqs. (\ref{smeared mpb}) and (\ref{smeared equilibrium density}) can be reduced to
\begin{flalign}
\left\{
1-(\gamma a)^2\nabla^2
\right\}\nabla^2\Psi({{\bf r}})=-4\pi l_B z^2\rho_{\mathrm{ref}}({\bf r})e^{-\Psi({\bf r})},
\label{real mpb1}
\end{flalign}
using eq. (\ref{gamma}) as the values of $\gamma$.
Thus, the higher-order PB equation of the BSK form [15-18] has been verified.

To validate another type of the higher-order PB equation given by eq. (\ref{higher-order pb-lue}), we need to perform the expansion of $f^{-1}({\bf k})$ up to the order $k^4$.
In the HNC approximation \cite{ng}, for example, it is straightforward to show that
\begin{equation}
\frac{1}{f ({\bf k})}=e^{\frac{(ka)^2}{4m}}\approx
1+\frac{1}{4m}(ka)^2+\frac{1}{32m^2}(ka)^4,
\label{gaussian for lue}
\end{equation}
yielding
\begin{flalign}
\nabla^2\left(1-\frac{a^2}{4m}\nabla^2+\frac{a^4}{32m^2}\nabla^4\right)\Psi({{\bf r}})
=-4\pi l_B z^2e^{\mu-J_1({\bf r})-\Psi ({\bf r})},
\label{our pb-lue}
\end{flalign}
with $m=1.08$.
In the parentheses on the left hand side of eq. (\ref{our pb-lue}), the coefficient of $\nabla^2$--term is close to that of of eq. (\ref{higher-order pb-lue}), and yet the coefficient of the $\nabla^4$--term is somewhat smaller than the previous factor in eq. (\ref{higher-order pb-lue}).

\section{Derivation scheme of the grand-potential density functional given by eqs. (\ref{omega sum}) to (\ref{v1 interaction energy})}
In this section, we outline how to derive eqs. (\ref{omega sum}) to (\ref{v1 interaction energy}) starting with the density functional integral representation of the grand potential.
More details of the following formulations are given in Appendix B.
\subsection{Density-functional integral representation of the grand potential for the attractive particle systems}
Let $\Omega[v,J]$ be the grand potential of non-uniform liquids where constituent particles in an external potential $J({\bf r})$ interact via an interaction potential $v({\bf r})$.
Once the density functional form of $\Omega[v,J]$ is found for attractive particle systems, other grand-potential density functionals for only repulsive systems (i.e., hard sphere fluids and point-charge systems) is obtained by setting that $v_{\mathrm{att}}({\bf r})\equiv 0$ and replacing $v_0$ with either hard sphere or Coulomb interaction potential (i.e., $v_{\mathrm{HS}}$ or $v_{\mathrm{el}}$).

As detailed in Appendix B, the grand potential can be represented by the functional integral over the density field as follows:
\begin{flalign}
e^{-\Omega[v,\,J]}
&=\int D\rho\,e^{-\Omega^*[\rho]}.
\label{dfi start}
\end{flalign}
Here the conditional grand potential $\Omega^*[\rho]$ of a given density field $\rho({\bf r})$ is an extension of the grand-potential density functional formulated in the conventional DFT [6-12]:
\begin{flalign}
\Omega^*[\rho]=\Omega_{\mathrm{V}}[\rho]+\Delta \Omega,
\label{conditional}
\end{flalign}
where $\Omega_{\mathrm{V}}[\rho]$ denotes the variational grand potential, which reduces to the equilibrium grand potential $\Omega$ when $\rho=\rho_{\mathrm{eq}}$, and $\Delta\Omega$ corresponds to the deviation free energy from the equilibrium grand potential due to the imposition of a given density field $\rho({\bf r})$, instead of $\rho_{\mathrm{eq}}({\bf r})$.
In this subsection, resulting forms of both $\Omega_{\mathrm{V}}[\rho]$ and $\Delta\Omega[\rho]$ are only presented (see Appendix B.1 for the detailed derivation).

The variational grand potential $\Omega_{\mathrm{V}}[\rho]$ is given by
\begin{flalign}
\Omega_{\mathrm{V}}[\rho]&=\mathcal{F}_0[\rho]-\int d{\bf r}\,\rho({\bf r})\mu+U_1[\rho],\label{frho}\\
U_1[\rho]&=\frac{1}{2}\iint d{\bf r}d{\bf r'}\rho({\bf r})v_{\mathrm{att}}({\bf r}-{\bf r}')\rho({\bf r}')+\int d{\bf r}\,\rho({\bf r})J_1({\bf r}).
\label{u1}
\end{flalign}
As expected from the subscript "0", the first term $\mathcal{F}_0[\rho]$ on the rhs of eq. (\ref{frho}) represents the intrinsic Helmholtz free energy of the only repulsive systems where particles in the absence of external field $J({\bf r})\equiv 0$ interact via the interaction potential $v_0({\bf r})$.
We can see from eq. (\ref{u1}) that the attractive interaction energy is evaluated in the mean-field approximation.
The second term on the rhs of eq. (\ref{u1}), which expresses the one-body potential energy, is coupled only with the slowly-varying external field $J_1$ because the configurational integration range is set outside the excluded region defined by eqs. (\ref{appendix radius}) and (\ref{appendix reference solute}).

Following the functional integral formulation combined with the conventional DFT, the first approximation of the additional contribution $\Delta\Omega$ provides the logarithmic form in the Gaussian approximation of the fluctuating one-body potential field, dual to a given density field $\rho({\bf r})$ (see Appendix B.2 for the derivation):
\begin{flalign}
\Delta \Omega&=\frac{1}{2}\ln\left|
\mathrm{det}\,G_0({\bf r};\rho^0_{\mathrm{ref}})
\right|
\label{delta f}\\
G_0({\bf r};\rho^0_{\mathrm{ref}})&=\rho^0_{\mathrm{ref}}\left\{
\delta({\bf r})+\rho^0_{\mathrm{ref}}h_0({\bf r})
\right\},
\label{g0 h}
\end{flalign}
where we have neglected variation of the density-density correlation function from $G_0({\bf r};\rho^0_{\mathrm{ref}})$ for a uniform density $\rho^0_{\mathrm{ref}}$, which is consistent with the other approximations of $\mathcal{F}_0[\rho]$ described below.

\subsection{Reference density $\rho_{\mathrm{ref}}$ as the saddle-point density}
Starting with the above density functional integral form (\ref{dfi start}), we would like to derive the mean-field density $\rho_{\mathrm{ref}}({\bf r})$ where the attractive part $v_{\mathrm{att}}({\bf r})$ of the interaction potential is considered.
To this end, we divide $\mathcal{F}_0[\rho]$ into the ideal gas and excess contributions, $\mathcal{F}_0^{\mathrm{id}}[\rho]$ and $\mathcal{F}_0^{\mathrm{ex}}[\rho]$:
\begin{equation}
\mathcal{F}_0[\rho]=\mathcal{F}_0^{\mathrm{id}}[\rho]+\mathcal{F}_0^{\mathrm{ex}}[\rho],
\label{gas excess}
\end{equation}
so that we may rewrite, for later convenience, the conditional grand potential $\Omega^*[\rho]$ given by eqs. (\ref{conditional}) to (\ref{g0 h}) as
\begin{flalign}
\Omega^*[\rho]&=\mathcal{F}_0^{\mathrm{ex}}[\rho]+\Omega_{\mathrm{att}}^*[\rho]
+\Delta\Omega,
\label{conditional2}\\
\Omega_{\mathrm{att}}^*[\rho]&=\mathcal{F}_0^{\mathrm{id}}[\rho]+U_1[\rho]
-\int d{\bf r}\,\rho({\bf r})\mu.
\label{attractive conditional}
\end{flalign}
Accordingly, eq. (\ref{dfi start}) can read
\begin{flalign}
e^{-\Omega[v,\,J]}
&=e^{-\Delta\Omega}\int D\rho\,e^{-\Omega_{\mathrm{att}}^*[\rho]}\left(
\frac{\int D\rho\,e^{-\mathcal{F}_0^{\mathrm{ex}}[\rho]-\Omega_{\mathrm{att}}^*[\rho]}}{\int D\rho\,e^{-\Omega_{\mathrm{att}}^*[\rho]}}
\right),
\label{dfi rewriting}
\end{flalign}
using eqs. (\ref{conditional2}) and (\ref{attractive conditional}).
This form (\ref{dfi rewriting}) reveals that the mean-field approximation of the attractive interaction energy becomes equivalent to the use of the saddle-point density determined by
\begin{equation}
\left.
\frac{\delta\Omega_{\mathrm{att}}^*[\rho]}{\delta\rho}\right|_{\rho=\rho_{\mathrm{ref}}}
=0
\label{sp attractive}
\end{equation}
because combination of eqs. (\ref{attractive conditional}) and (\ref{sp attractive}) yields
\begin{flalign}
\ln\rho_{\mathrm{ref}}({\bf r})
+\int d{\bf r}'v_{\mathrm{att}}({\bf r}-{\bf r}')\rho_{\mathrm{ref}}({\bf r}')
+J_1({\bf r})-\mu=0,
\label{our attractive scf}
\end{flalign}
which is identical to the previous result (\ref{mean-field reference}).

\subsection{Hybrid evaluation based on the mean-field and Gaussian approximations}
Following the previous treatment, we approximate the functional integral (\ref{dfi rewriting}) by
\begin{flalign}
e^{-\Omega[v,\,J]}
\approx
e^{-\widetilde{\Omega}[\rho_{\mathrm{ref}}]}
\equiv
e^{-\Omega_{\mathrm{att}}^*[\rho_{\mathrm{ref}}]-\Delta\Omega}\left(
\frac{\int Dn\,e^{-\mathcal{F}_0[\rho_{\mathrm{ref}}+n]-U_1[\rho_{\mathrm{ref}}+n]}}{e^{-\Omega_{\mathrm{att}}^*[\rho_{\mathrm{ref}}]}}
\right),
\label{dfi rewriting2}
\end{flalign}
extracting only the saddle-point path for the attractive part of the grand potential.
The approximate form (\ref{dfi rewriting2}) of $\Omega[v,\,J]$ has been referred to as the grand-potential density functional which is denoted by $\widetilde{\Omega}[\rho_{\mathrm{ref}}]$ in eq. (\ref{omega sum}).
We expand $\mathcal{F}_0[\rho_{\mathrm{ref}}+n]+U_1[\rho_{\mathrm{ref}}+n]$ around the reference energy up to the quadratic term with respect to fluctuating density field $n({\bf r})$, yielding (see Appendix B.2 for the derivation)
\begin{flalign}
&\mathcal{F}_0[\rho_{\mathrm{ref}}+n]+U_1[\rho_{\mathrm{ref}}+n]-\Omega_{\mathrm{att}}^*[\rho_{\mathrm{ref}}]\nonumber\\
&\quad\approx\mathcal{F}^{\mathrm{ex}}_0[\rho_{\mathrm{ref}}]
-\iint d{\bf r}d{\bf r}'
\,n({\bf r})\,c_0({\bf r}-{\bf r}')\Delta\rho_{\mathrm{ref}}({\bf r}')\nonumber\\
&\qquad\qquad+\frac{1}{2}\iint d{\bf r}d{\bf r}'\,
n({\bf r})\,G_0^{-1}({\bf r}-{\bf r}';\rho_{\mathrm{ref}}^0)\,n({\bf r}'),
\label{expansion4integral}
\end{flalign}
supposing that the intrinsic Helmholtz free energy $\mathcal{F}_0[\rho_{\mathrm{ref}}]$ for the $v_0$-interaction systems are of the Ramakrishnan-Yussouff functional form [1, 6, 7]:
\begin{flalign}
\mathcal{F}_0[\rho_{\mathrm{ref}}]=\mathcal{F}_0[\rho^0_{\mathrm{ref}}]-\frac{1}{2}\iint d{\bf r}d{\bf r'}\Delta\rho_{\mathrm{ref}}({\bf r})c_0({\bf r}-{\bf r}')\Delta\rho_{\mathrm{ref}}({\bf r}')+\int d{\bf r}\rho_{\mathrm{ref}}({\bf r})\ln\left\{
\frac{\rho_{\mathrm{ref}}({\bf r})}{\rho^0_{\mathrm{ref}}}
\right\},
\label{our ry}
\end{flalign}
where the same approximation used in eqs. (\ref{underline density}), (\ref{delta f}), and (\ref{g0 h}) has been applied to the last line of eq. (\ref{expansion4integral}) (see also eq. (\ref{appendix g0 approx})).

Accordingly, eq. (\ref{dfi rewriting2}) is further reduced to
\begin{flalign}
e^{-\widetilde{\Omega}[\rho_{\mathrm{ref}}]}
&=
e^{-\mathcal{F}_0^{\mathrm{ex}}[\rho_{\mathrm{ref}}]-\Omega_{\mathrm{att}}^*[\rho_{\mathrm{ref}}]-\Delta\Omega}
\left(
\int Dn\,e^{-\Delta \mathcal{F}_0[n]}
\right)
\label{dfi rewriting3}\\
\Delta \mathcal{F}_0[n]
&=\frac{1}{2}\iint d{\bf r}d{\bf r}'\,
n({\bf r})\,G_0^{-1}({\bf r}-{\bf r}';\rho_{\mathrm{ref}}^0)n({\bf r}')
-\iint d{\bf r}d{\bf r}'
\,n({\bf r})c_0({\bf r}-{\bf r}')\Delta\rho_{\mathrm{ref}}({\bf r}'),
\label{delta u0}
\end{flalign}
indicating that the remaining task is to perform the functional integral over the fluctuating $n$-field.
As proved in Appendix B.3, eq. (\ref{delta u0}) reads
\begin{flalign}
\Delta \mathcal{F}_0[n]
&=\frac{1}{2}\iint d{\bf r}d{\bf r}'\,
\widetilde{n}({\bf r})\,G_0^{-1}({\bf r}-{\bf r}';\rho_{\mathrm{ref}}^0)\widetilde{n}({\bf r}')
\nonumber\\
&\quad-\frac{1}{2}\iint d{\bf r}d{\bf r}'\Delta\rho_{\mathrm{ref}}({\bf r})\left\{
h_0({\bf r}-{\bf r}')-c_0({\bf r}-{\bf r}')
\right\}\Delta\rho_{\mathrm{ref}}({\bf r}'),
\label{shifted delta u0}
\end{flalign}
due to the shift of fluctuating field from $n$ to
\begin{flalign}
\widetilde{n}({\bf r})&=n({\bf r})
-\iint d{\bf r}'d{\bf r}"\>G_0({\bf r}-{\bf r}';\rho_{\mathrm{ref}}^0)c_0({\bf r}'-{\bf r}")\Delta\rho_{\mathrm{ref}}({\bf r}")\nonumber\\
&=n({\bf r})-\iint d{\bf r}'d{\bf r}"\>\rho_{\mathrm{ref}}^0\left\{
\delta({\bf r}-{\bf r}')+\rho_{\mathrm{ref}}^0h_0({\bf r}-{\bf r}'))\right\}
c_0({\bf r}'-{\bf r}")\Delta\rho_{\mathrm{ref}}({\bf r}")\nonumber\\
&=n({\bf r})-\rho_{\mathrm{ref}}^0\int d{\bf r}"h_0({\bf r}-{\bf r}")\Delta\rho_{\mathrm{ref}}({\bf r}")\nonumber\\
&\approx
n({\bf r})-\left\{\rho_{\mathrm{eq}}({\bf r})-\rho_{\mathrm{ref}}({\bf r})
\right\},
\label{shift n}
\end{flalign}
where the last approximation is the same as eq. (\ref{underline density}).
It follows from eqs. (\ref{dfi rewriting3}) and (\ref{shifted delta u0}) that the Gaussian integration over the $\widetilde{n}$--field leads to
\begin{flalign}
e^{-\widetilde{\Omega}[\rho_{\mathrm{ref}}]}
&=e^{-\mathcal{F}^{\mathrm{ex}}_0[\rho_{\mathrm{ref}}]-\Omega_{\mathrm{att}}^*[\rho_{\mathrm{ref}}]-\Delta\Omega+\frac{1}{2}\iint d{\bf r}d{\bf r}'\Delta\rho_{\mathrm{ref}}({\bf r})\left\{
h_0({\bf r}-{\bf r}')-c_0({\bf r}-{\bf r}')\right\}\Delta\rho_{\mathrm{ref}}({\bf r}')}\nonumber\\
&\qquad\qquad\qquad\qquad\qquad\qquad
\times\underline{\left\{
\int D\widetilde{n}\,e^{-\frac{1}{2}\iint d{\bf r}d{\bf r}'\widetilde{n}({\bf r})G_0^{-1}({\bf r}-{\bf r}')
\widetilde{n}({\bf r}')}
\right\}}
\nonumber\\
&=e^{-\mathcal{F}^{\mathrm{ex}}_0[\rho_{\mathrm{ref}}]-\Omega_{\mathrm{att}}^*[\rho_{\mathrm{ref}}]+\frac{1}{2}\iint d{\bf r}d{\bf r}'\Delta\rho_{\mathrm{ref}}({\bf r})\left\{
h_0({\bf r}-{\bf r}')-c_0({\bf r}-{\bf r}')\right\}\Delta\rho_{\mathrm{ref}}({\bf r}')},
\label{gaussian integral}
\end{flalign}
where the logarithmic term $\Delta\Omega$ given by eq. (\ref{delta f}) is canceled by the result of Gaussian integration over the $\widetilde{n}$--field in the underlined term.
Combining eqs. (\ref{attractive conditional}), (\ref{our ry}) and (\ref{gaussian integral}), we obtain eqs. (\ref{omega sum}) to (\ref{v1 interaction energy}), the central results for $\widetilde{\Omega}[\rho_{\mathrm{ref}}]$.

\section{Summary and conclusions}
\begin{figure}[t]
\begin{center}
	\includegraphics[
	width=15.5 cm
]{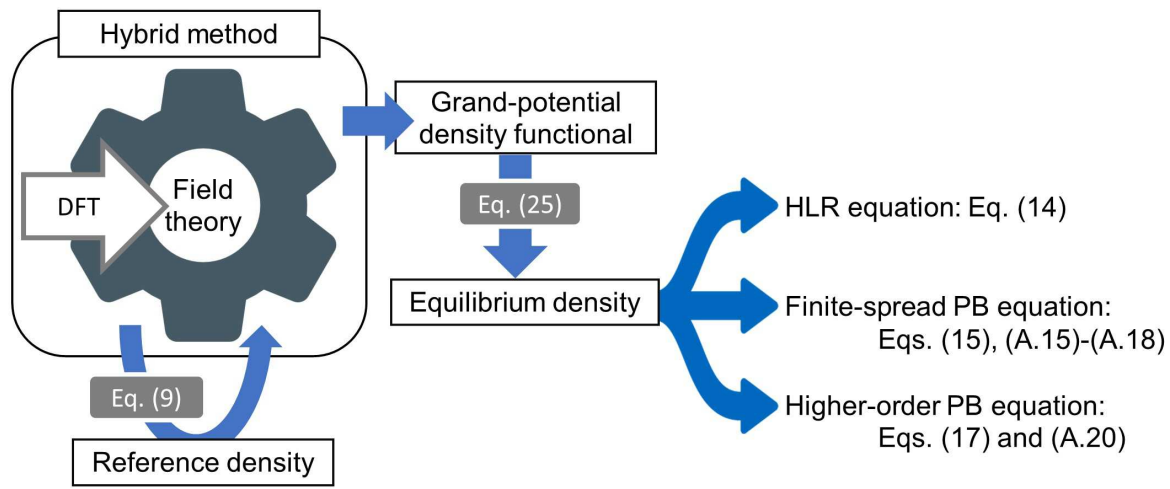}
\end{center}
\caption{
A schematic flow chart of proving the equivalence of previous SC equations [4, 5, 14-34, 42-47] based on the hybrid method \cite{frusawa2,frusawa3} that incorporates the density functional theory (DFT) [6-12] into the statistical field theory [15, 23, 31-34, 36, 50-52]. The saddle-point equation (\ref{mean-field reference}), or eq. (\ref{sp attractive}), determines the reference density $\rho_{\mathrm{ref}}({\bf r})$. The Gaussian approximation of density fluctuations around $\rho_{\mathrm{ref}}({\bf r})$ yields the grand-potential density functional $\widetilde{\Omega}[\rho_{\mathrm{ref}}]$ given by eqs. (\ref{omega sum}) to (\ref{v1 interaction energy}). The equilibrium density $\rho_{\mathrm{eq}}({\bf r})$ is obtained from $\widetilde{\Omega}[\rho_{\mathrm{ref}}]$ using the relation (\ref{general equilibrium density}). The single unified equation (\ref{tcf equilibrium density}) for $\rho_{\mathrm{eq}}({\bf r})$ produces a variety of previous SC equations, including the hydrostatic linear response (HLR) equation (\ref{weeks hlr}) [4, 5, 14, 42-47], the finite-spread Poisson-Boltzmann (PB) equation (\ref{weeks PB}) (or eqs. (A.15) to (A.18)) [14, 20-24], and the higher-order PB equation (\ref{higher-order pb-lue}) (or eq. (A.20)) [15-18, 25-30].
}
\end{figure}
The flow chart depicted in Fig. 1 outlines the unified description of previous SC equations covering the mean-field equation (\ref{mean-field reference}), the HLR equation (\ref{weeks hlr}) [4, 5, 14, 42-47], the finite-spread PB equation [14, 20-24] given by eq. (\ref{weeks PB}), and the higher-order PB equation [15-18, 25-30] expressed by eq. (\ref{higher-order pb-lue}).
The first step is that the hybrid method \cite{frusawa2,frusawa3}, as a kind of thermodynamic perturbation theory, yields the grand-potential density functional $\widetilde{\Omega}[\rho_{\mathrm{ref}}]$ of the reference density $\rho_{\mathrm{ref}}({\bf r})$ as a solution of the saddle-point equation (\ref{sp attractive}), or eq. (\ref{mean-field reference}).
Next, a single unified equation (\ref{tcf equilibrium density}) for $\rho_{\mathrm{eq}}({\bf r})$ is obtained from $\widetilde{\Omega}[\rho_{\mathrm{ref}}]$ using the relation (\ref{general equilibrium density}).
The final step is to demonstrate that the unified equation (\ref{tcf equilibrium density}) is reduced to the three types of SC equations [4, 5, 14-34, 42-47].

Following the process in Fig. 1, we can divide the formulations presented thus far into the following four parts: (i) the mean-field equation (\ref{mean-field reference}) for the reference density $\rho_{\mathrm{ref}}({\bf r})$, (ii) the grand-potential density functional $\widetilde{\Omega}[\rho_{\mathrm{ref}}]$ given by eqs. (\ref{omega sum}) to (\ref{v1 interaction energy}), (iii) the HLR equation (\ref{weeks hlr}) for the equilibrium density $\rho_{\mathrm{eq}}({\bf r})$ [4, 5, 14, 42-47], and (iv) the modified PB equations (\ref{smeared mpb}), (\ref{smeared equilibrium density}), (\ref{real mpb1}), and (\ref{our pb-lue}) [14-30].
A detailed summary of each part is provided below.

{\itshape (i) The mean-field equation (\ref{mean-field reference}) for the reference density $\rho_{\mathrm{ref}}({\bf r})$}.---
The proposed formulation clarifies that the HLR theory [4, 5, 14, 42-47] proceeds in the opposite direction to the conventional thermodynamic perturbation [1-3] theory that treats the slowly varying attractive interaction potential as a perturbation to the repulsive reference system.
In contrast to the conventional thermodynamic perturbation theory, our formulations based on the HLR treatment adopt, as a reference free energy, the mean-field free energy of attractive particle system under an external field in the absence of a repulsive interaction potential, i.e. $v_0({\bf r})\equiv 0$.
It is the aim of the saddle-point equation (\ref{sp attractive}) for $\rho_{\mathrm{ref}}({\bf r})$ to determine a reference state that reflects the entire profile of the non-uniform density distribution using a simple and precise method.
Along with the HLR treatment, we investigated a perturbative contribution to the reference system from energetic aspects when recovering the repulsive part $v_0({\bf r})$ of the interaction potential.

{\itshape (ii) The grand-potential density functional $\widetilde{\Omega}[\rho_{\mathrm{ref}}]$}.---
The hybrid method leads to a novel density-functional form of the grand potential given by eqs. (\ref{omega sum}) to (\ref{v1 interaction energy}), thereby demonstrating that the same density-functional form validates the previous SC equations [4, 5, 14-30, 42-47].
It has been shown that there are two main features of the density-functional form.
One is that $\widetilde{\Omega}[\rho_{\mathrm{ref}}]$ is a functional of the reference density evaluated in the mean-field approximation, instead of the equilibrium density $\rho_{\mathrm{eq}}({\bf r})$ used in the conventional DFT.
The other is that the TCF, specifically $-h_0({\bf r})$, is treated as effective interaction potential instead of the DCF.
The TCF, which is generated from the DCF due to the Ornstein-Zernike equation, emerges as a result of considering fluctuations around $\rho_{\mathrm{ref}}({\bf r})$ when repulsive interactions are inserted into the systems as described above.
In more detail, the second term on the rhs of eq. (\ref{expansion4integral}), which is the coupling term for the fluctuating density $n({\bf r})$ with the reference density $\rho_{\mathrm{ref}}({\bf r})$, plays a key role in generating the effective interaction energy expressed by the TCF.
The non-vanishing coupling term reveals that the proposed formulation follows the core concept of the HLR theory [4, 5, 14, 42-47]. 

{\itshape (iii) The HLR equation (\ref{weeks hlr}) for the equilibrium density $\rho_{\mathrm{eq}}({\bf r})$}.---
The difference between the HLR method and our theory is that we treat the slowly varying part of the repulsive interaction potential, including the long-range part of the Coulomb interaction potential, as a perturbative potential added in the reference system, even though the other slowly varying parts are evaluated in the mean-field approximation according to the previous SC equations reviewed in Section 2.
The local molecular field theory [14, 20, 21] starts with the {\itshape a priori} separation of $v_{\mathrm{el}}({\bf r})$ into harshly repulsive and slowly varying parts, thus providing the finite-spread PB equation [14, 20-24] or the aforementioned mean-field equation based on a Gaussian charge smearing model with the harshly repulsive part of $v_{\mathrm{el}}({\bf r})$ to be cut-off [14, 20-24, 35-40]. 
However, our goal is to clarify the underlying mechanisms that generate a variety of SC equations from a unified perspective.
Therefore, we investigated the theoretical background of why the Gaussian charge smearing model is relevant for describing charge-charge correlations.

As a result, we verified the HLR equation [4, 5, 14, 42-47] for any repulsive interaction system using a unified framework based on the recently developed density functional integral representation [48, 49].
The field-theoretic formulation of the HLR equation also demonstrates the benefits of the hydrostatic approximation [4, 5, 14, 42-47], which regards a uniform liquid with a density of $\rho_{\mathrm{ref}}^0$, instead of the bulk density, as another reference system for evaluating the grand-potential density functional of non-uniform liquids.

{\itshape (iv) A set of modified PB equations, eqs. (\ref{smeared mpb}), (\ref{smeared equilibrium density}), (\ref{real mpb1}) and (\ref{our pb-lue})}.---
In terms of the Coulomb potential $\Psi({\bf r})$ defined as eq. (\ref{equilibrium el potential}) using the TCF, we can see that the HLR equation (\ref{weeks hlr}), expressed by the TCF, forms the basis of validating the Boltzmann distribution (\ref{equilibrium el density}) for $\rho_{\mathrm{eq}}({\bf r})$ in both the finite-spread PB equation [14, 20-24] and higher-order PB equation [15-18, 25-30].
These modifications of the PB equation are ascribed to the Poisson equation (\ref{dcf poisson}) of the electrostatic DCF $c_{\mathrm{el}}({\bf r})$.
Furthermore, the Ornstein-Zernike equation yields the SC equation (\ref{oz potential}) for $\Psi({\bf r})$ with the use of $c_{\mathrm{el}}({\bf r})$.
Combining these relations, eqs. (\ref{oz potential}) and (\ref{dcf poisson}), we obtain the linearized SC equation (\ref{linear pb}) for $\Psi({\bf r})$, which transforms to alternative representations, the finite and higher-order PB equations, depending on how the weight function $f({\bf r})$ is approximated and/or interpreted.
We stress again that these modifications of the conventional PB equation emerge due to the electrostatic DCF, and that the electrostatic DCF naturally appears when calculating $\rho_{\mathrm{eq}}({\bf r})$ from the grand-potential density functional $\widetilde{\Omega}[\rho_{\mathrm{ref}}]$.

In conclusion, we have proved the equivalence of previous SC equations [4, 5, 14-34, 42-47] by demonstrating that the apparent diversity of SC equations can be explained by the identical form of the grand-potential density functional given by eqs. (\ref{omega sum}) to (\ref{v1 interaction energy}).
This theoretical finding implies that not only improvements to the previous forms but also derivations of any new equations can be performed in a systematic manner, by either exploring a new system to which our formulations may apply or considering higher-order terms beyond the Gaussian approximation.

\appendix
\section{Summary of previous results related to modified PB equations for point-charge systems}
\subsection{Coupling constant in non-uniform systems}
Let $a$ and $\Gamma$ be the mean distance between point charges and the coupling constant representing the strength of Coulomb interactions, respectively.
As is well-known, $\Gamma$ in the uniform OCP is given by
\begin{flalign}
\Gamma=z^2\frac{l_B}{a},
\label{appendix gamma}
\end{flalign}
when the valence of each charge is $z$.
In non-uniform point-charge systems, however, the distance between charges cannot be uniquely determined.
Correspondingly, it is not trivial to define $\Gamma$, unlike the conventional uniform OCP.
Two types of coupling constants have been defined so far for the typical non-uniform systems of small counterions that are dissociated from oppositely charged macroions, or macroscopic objects including charged plates:
while one uses the Gouy-Chapman length as a reference length $a$, the other definition of $\Gamma$ is based on an analogy with the two-dimensional OCP.

We have adopted the latter definition as follows.
When all of the conterions are localized on the entire surface of macroions oppositely charged, the distance $b$ between counterions on the surface is defined by
\begin{equation}
\pi b^2\sigma=z,
\label{appendix 2D distance}
\end{equation}
using the number density $\sigma$ per area of surface charges.
Obviously, $b$ is the minimum of counterion-counterion separation: $b\leq a$.
The second definition of $\Gamma$ in counterion systems therefore implies that eq. (\ref{appendix gamma}) leads to the unique definition of $\Gamma$ for non-uniform systems as long as $a$ is identified with $b$.
That is, we set that $a=b$ not only in eq. (\ref{appendix gamma}) but also in the main text.
Consequently, the relation $\Gamma\gg 1$ continues to represent the strong charge-charge correlations in non-uniform systems.

\subsection{Two types of weighted densities}
There are two methods to formulate charge smearing models [14, 20-24, 35-40] although not described explicitly.
To see this, we first investigate the relationship between three density functional forms of the slowly varying part of the Coulomb interaction energy $U_1^{\mathrm{el}}$ that is associated with the slowly varying part of the Coulomb interaction potential such as eq. (\ref{Coulomb v1}).

The first form expresses $U_1^{\mathrm{el}}$ as the sum of Coulomb interactions between smeared charges with the interaction potential being given by the original interaction potential $v_{\mathrm{el}}({\bf r})$:
\begin{flalign}
U_1^{\mathrm{el}}[\rho]
&=\frac{z^2l_B}{2}\iint d{\bf r}d{\bf r}'\,
\overline{\,\rho}({\bf r})v^{\mathrm{el}}({\bf r}-{\bf r}')
\,\overline{\,\rho}({\bf r}'),
\label{appendix smeared energy1}\\
\overline{\,\rho}({\bf r})
&=\int d{\bf s}\,\rho({\bf s})\,\omega({\bf s}-{\bf r}),
\label{appendix smeared density1}
\end{flalign}
where $\overline{\,\rho}({\bf r})$ is smeared due to the weighted function $\omega({\bf s}-{\bf r})$;
for instance, we have
\begin{flalign}
\omega({\bf r})=
\left\{
\begin{array}{l}
\frac{1}{\{\pi a^2/(2m^2)\}^{3/2}}e^{-2\left(\frac{mr}{a}\right)^2}\quad(\mathrm{Gaussian})\\
\\
\frac{3}{4\pi a^3}\Theta(a-r)\quad\>(\mathrm{Homogeneous})\\
\\ 
\frac{2}{\pi a^2}\left(
\frac{K_1(2r/a)}{r}
\right)\quad\>\>\>\>\>(\mathrm{Bessel}),
\end{array}
\right.
\label{appendix omega forms}
\end{flalign}
which are the models that consider the radial distribution given by the modified Bessel function $K_1$ [26] as well as the Gaussian and homogeneous charge distributions.

Equations (\ref{appendix smeared energy1}) and (\ref{appendix smeared density1}) naturally lead to the second description of $U_1^{\mathrm{el}}$ where the slowly varying interaction potential $v_1({\bf r})$ is introduced without smearing the density:
\begin{flalign}
U_1^{\mathrm{el}}[\rho]
&=\frac{z^2l_B}{2}\iint d{\bf s}d{\bf s}'
\rho({\bf s})v_1^{\mathrm{el}}({\bf s}-{\bf s}')\rho({\bf s}').
\label{appendix smeared energy2}
\end{flalign}
Comparing eqs. (\ref{appendix smeared energy1}), (\ref{appendix smeared density1}), and (\ref{appendix smeared energy2}), it is found that
\begin{flalign}
v_1^{\mathrm{el}}({\bf s}-{\bf s}')
&=\iint d{\bf r}d{\bf r}'\,\omega({\bf s}-{\bf r})\,
v^{\mathrm{el}}({\bf r}-{\bf r}')\,
\omega({\bf r}'-{\bf s}'),
\label{appendix smeared interaction1}
\end{flalign}
which also reads
\begin{flalign}
v_1^{\mathrm{el}}({\bf s}-{\bf s}')
=\iint d{\bf r}'d{\bf r}"\,
v^{\mathrm{el}}({\bf s}-{\bf r}")\,\omega({\bf r}"-{\bf r}')\,\omega({\bf r}'-{\bf s}')
\label{appendix smeared interaction2}
\end{flalign}
due to the coordinate change such that ${\bf r}"={\bf s}-{\bf r}+{\bf r}'$.

Equations (\ref{appendix smeared energy2}) and (\ref{appendix smeared interaction2}) thus generate the third form:
\begin{flalign}
U_1^{\mathrm{el}}[\rho]
&=\frac{z^2l_B}{2}\iint d{\bf r}d{\bf r}"
\rho({\bf r})v^{\mathrm{el}}({\bf r}-{\bf r}")\overline{\overline{\,\rho}}({\bf r}"),
\label{appendix smeared energy3}\\
\overline{\overline{\,\rho}}({\bf r}")
&=\iint d{\bf r}'d{\bf s}'\,\omega({\bf r}"-{\bf r}')\,\omega({\bf r}'-{\bf s}')\rho({\bf s}').
\label{appendix smeared density3}
\end{flalign}
Equation (\ref{appendix smeared density3}) represents the second type of the weighted density used in the main text;
as will be seen below, the expression (\ref{appendix smeared density3}) yields the same form as the distribution function given in eq. (\ref{weeks gaussian}).

\subsection{Finite-spread PB equations [14, 20-24]}
There are two kinds of expressions for the Coulomb potentials created by counterion-counterion interactions in correspondence with the above two definitions of smeared density, i.e., $\overline{\,\rho_{\mathrm{ref}}}({\bf r})$ and $\overline{\overline{\,\rho_{\mathrm{ref}}}}({\bf r})$.
While a finite-spread PB theory [22] uses the former definition,
\begin{flalign}
\psi({\bf r})
=\int d{\bf r}'v^{\mathrm{el}}({\bf r}-{\bf r}')\,\overline{\,\rho_{\mathrm{ref}}}({\bf r}'),
\label{appendix  frydel potential}
\end{flalign}
the local molecular field theory [14, 20, 21], another finite-spread PB theory, adopts the latter:
\begin{flalign}
\Psi ({\bf r})
&=\int d{\bf r}'\,v^{\mathrm{el}}({\bf r}-{\bf r}')\overline{\overline{\,\rho_{\mathrm{ref}}}}({\bf r}')
\label{appendix weeks potential1}\\
&=\int d{\bf r}'\,v_1^{\mathrm{el}}({\bf r}-{\bf r}')\rho_{\mathrm{ref}}({\bf r}').
\label{appendix weeks potential2}
\end{flalign}
Both potentials are related to each other as follows:
\begin{equation}
\Psi ({\bf r})=\int d{\bf r}'\,\omega({\bf r}-{\bf r}')\psi({\bf r}'),
\label{appendix two potentials}
\end{equation}
indicating that $\Psi ({\bf r})$ corresponds to the weighted potential of $\psi({\bf r})$.

We can see from eq. (\ref{appendix weeks potential2}), the key representation, that the reference density given by eq. (\ref{mean-field reference})  obeys the Boltzmann distribution such that
\begin{flalign}
\rho_{\mathrm{ref}}({\bf r})
&=e^{\mu-J_1({\bf r})-\int d{\bf r}'\,\omega({\bf r}-{\bf r}')\psi({\bf r}')}
\label{appendix boltzmann1}\\
&=e^{\mu-J_1({\bf r})-\Psi({\bf r})}.
\label{appendix boltzmann2}
\end{flalign}
Furthermore, it is found from the above definitions given by eqs. (\ref{appendix weeks potential1}) and (\ref{appendix weeks potential2}) that the Poisson equations of $\psi ({\bf r})$ and $\Psi ({\bf r})$ yield
\begin{flalign}
\nabla^2\psi ({\bf r})
&=-4\pi z^2l_B\int d{\bf r}'\,\omega({\bf r}-{\bf r}')\,\rho_{\mathrm{ref}}({\bf r}'),
\label{appendix poisson1}\\
\nabla^2\Psi ({\bf r})
&=-4\pi z^2l_B\iint d{\bf r}'d{\bf r}"\,\omega({\bf r}-{\bf r}')\,\omega({\bf r}'-{\bf r}")\rho_{\mathrm{ref}}({\bf r}"),
\label{appendix poisson2}
\end{flalign}
respectively.
Combining eqs. (\ref{appendix boltzmann1}) and (\ref{appendix poisson1}), we obtain an expression of the finite-spread PB equation.
As mentioned at the end of Appendix A.3, on the other hand, eq. (\ref{appendix poisson2}) reads
\begin{flalign}
\nabla^2\Psi ({\bf r})
&=-4\pi z^2l_B\left\{
\frac{8}{\{\pi a^2/m^2\}^{3}}\iint d{\bf r}'d{\bf r}"\,e^{-2|m({\bf r}-{\bf r}')/a|^2}\,e^{-2|m({\bf r}'-{\bf r}")/a|^2}\rho_{\mathrm{ref}}({\bf r}")\right\}
\nonumber\\
&=-4\pi z^2l_B\left\{
\frac{1}{\{\pi a^2/m^2\}^{3/2}}\int d{\bf r}"\,e^{-|m({\bf r}-{\bf r}")/a|^2}\rho_{\mathrm{ref}}({\bf r}")\right\}
\label{appendix poisson-gaussian}
\end{flalign}
for Gaussian charge smearing model.
Combining eqs. (\ref{appendix boltzmann2}) and (\ref{appendix poisson-gaussian}), we verify another form of the finite-spread PB equation given by eqs. (\ref{weeks PB}) and (\ref{weeks gaussian}) used in the local molecular field theory [14, 20, 21].

\subsection{Higher-order PB equations [15-18, 25-30]}
One type of higher-order PB equations is truncated at the biharmonic $\nabla^2\nabla^2$-term, which can be written as
\begin{flalign}
\left\{(\gamma a)^2\nabla^2-1\right\}\nabla^2\Psi({{\bf r}})
=4\pi l_B z^2\rho_{\mathrm{ref}}({\bf r})e^{-\Psi({\bf r})},
\label{appendix another higher-order pb}
\end{flalign}
where $\gamma$ takes the value of either $\Gamma$ or $\sim 10^0$ [16-18, 25, 28].
This type of equation is the same as the BSK equation in binary systems as mentioned in the main text.
The BSK equation [16-18] has been found to be a powerful tool in investigating the structural and dynamical properties of concentrated electrolytes and room temperature ionic liquids where $a$ is identified with the diameter of ions.

Meanwhile, the other type has been presented in eq (\ref{higher-order pb-lue}) which has the $\nabla^2\nabla^4$-term as the gradient term of the highest order.
Incidentally, in eq. (\ref{higher-order pb-lue}), the coefficients of the $\nabla^2\nabla^2$--and $\nabla^2\nabla^4$--terms use a previous result of $\sigma$, given that $\Delta=0$ in eq. (26) of Ref \cite{lue}.

\subsection{Correspondence between the electrostatic DCF $c_{\mathrm{el}}({\bf r})$ and the slowly varying part $v_1^{\mathrm{el}}({\bf r})$ of the Coulomb interaction potential}
The electrostatic DCF relevant in the strong coupling regime of the OCP can be regarded as $v_1^{\mathrm{el}}({\bf r})$ when considering specific charge smearing models.
Comparison between eqs. (\ref{dcf el}) and (\ref{appendix smeared interaction2}) implies that the equality,
\begin{equation}
f({\bf r}-{\bf r}')=\int d{\bf r}"\omega({\bf r}-{\bf r}")\,\omega({\bf r}"-{\bf r}'),
\label{appendix f omega}
\end{equation}
holds between $f({\bf r})$ and $\omega({\bf r})$ for the strongly coupled OCP.
More specifically, $f({\bf r})$ in the HNC approximation \cite{ng} is expressed by $\omega({\bf r})$ of the Gaussian charge smearing model, whereas $f({\bf r})$ in the soft MSA \cite{frydel2016,rosenfeld gelbart,rosenfeld evans} by $\omega({\bf r})$ of the homogeneous charge smearing model (see also eqs. (\ref{form factor}) and (\ref{appendix omega forms})).

\section{Details of Section 6}
\subsection{Density functional form of the conditional grand potential $\Omega^*[\rho]$: derivation of eqs. (\ref{dfi start}) to (\ref{u1}) \cite{frusawa2,frusawa3}}
The configurational representation $\Omega[v,\,J]$ is
\begin{align}
&e^{-\Omega[v,\,J]}=\mathrm{Tr}\,\exp\left\{
-\sum_{i<j}v({\bf r}_i-{\bf r}_j)
-\sum_{i}J({\bf r}_i)
\right\}\nonumber\\
&\mathrm{Tr}\equiv
\sum_{N=0}^{\infty}e^{N\mu}\frac{1}{N!}\int d{\bf r}_1\cdots\int d{\bf r}_N,
\label{appendix omega}
\end{align}
where the interaction energy $\sum_{i<j}v({\bf r}_i-{\bf r}_j)$ appearing in the above exponent can read, according to the potential splitting given by eq. (\ref{interaction split}),
\begin{flalign}
\sum_{i<j}v({\bf r}_i-{\bf r}_j)&=\sum_{i<j}v_0({\bf r}_i-{\bf r}_j)+\frac{1}{2}\iint d{\bf r}d{\bf r'}\hat{\rho}({\bf r})v_{\mathrm{att}}({\bf r}-{\bf r}')\hat{\rho}({\bf r}')
\label{appendix split}
\end{flalign}
where $\hat{\rho}({\bf r})=\sum_{i=1}^N\delta({\bf r}-{\bf r}_i)$ represents an instantaneous density of $N$ spheres located at $\{{\bf r}_i\}\,(i=1,\cdots ,N)$.
Bearing in mind the expression (\ref{appendix split}), the density-functional integral representation can be incorporated into eq. (\ref{appendix omega}) via the following relation:
\begin{flalign}
e^{-\Omega[v,\,J]}&=\int D\rho\,\prod_{{\bf r}}\delta\left[
\hat{\rho}({\bf r})-\rho({\bf r})
\right]e^{-\Omega[v_0, 0]-U_1[\rho]}\nonumber\\
&=\int D\rho\left(
\int D\psi\,e^{\int d{\bf r}\,\,i\psi({\bf r})\{\hat{\rho}({\bf r})-\rho({\bf r})\}}e^{-\Omega[v_0,0]-U_1[\rho]}
\right)\nonumber\\
&=\int D\rho\,e^{-\Omega^*[\rho]},
\label{appendix dfi start}\\
U_1[\rho]&=\frac{1}{2}\iint d{\bf r}d{\bf r'}\rho({\bf r})v_{\mathrm{att}}({\bf r}-{\bf r}')\rho({\bf r}')
+\int d{\bf r}\,\rho({\bf r})J({\bf r}),
\label{appendix u1}
\end{flalign}
where $\prod_{{\bf r}}\delta\left[\hat{\rho}({\bf r})-\rho({\bf r})\right]$ represents the constraint that a density field $\rho({\bf r})$ must be the same as an instantaneous density distribution $\hat{\rho}({\bf r})$ at every position in the system and an identity, $1=\int D\rho\,\prod_{{\bf r}}\delta\left[\hat{\rho}({\bf r})-\rho({\bf r})\right]$, has been inserted into the first line of the above equations.

The purpose of this subsection is to find the density-functional form of the conditional grand potential $\Omega^*[\rho]$ with the help of the conventional DFT [6-12].
To this end, it is necessary to manipulate the grand potential $\Omega[v_0,-i\psi]$ under the complex external field $-i\psi({\bf r})$ because eq. (\ref{appendix dfi start}) is converted to
\begin{flalign}
e^{-\Omega^*[\rho]}&=
\int D\psi\,
e^{-U_1[\rho]-i\int d{\bf r}\,\rho({\bf r})\psi({\bf r})}
\,\mathrm{Tr}\,
\,e^{-\sum_{i<j}v_0({\bf r}_i-{\bf r}_j)
+\sum_{i}i\psi({\bf r}_i)}\nonumber\\
&=e^{-U_1[\rho]}\,
\underline{\left(\int D\psi\,
e^{-\Omega[v_0,-i\psi]-i\int d{\bf r}\,\rho({\bf r})\psi({\bf r})}\right)},
\label{appendix conditional free}
\end{flalign}
implying that we can obtain the density functional from evaluating the underlined $\psi$--integral.

Let the complex potential field $\psi$ be decomposed into two parts of the fluctuating field $\phi$ and the saddle-point field $i\psi^*$: $\psi=\phi+i\psi^*$.
On the one hand, the saddle-point path provides the first Legendre transform of $\Omega[v_0,-i\psi]$: 
\begin{align}
&\left.
\frac{\delta\left(\Omega[v_0,-i\psi]\right)}{\delta\psi({\bf r})}\right|_{\psi=i\psi^*}
=-i\rho({\bf r}),
\label{appendix omega psi}\\
&\mathcal{F}_0[\rho]
=\Omega[v_0,\psi^*]-\int d{\bf r}\,\rho({\bf r})\left\{\psi^*({\bf r})-\mu\right\},
\label{appendix legendre}
\end{align}
where $\mathcal{F}_0[\rho]$ corresponds to the {\itshape intrinsic} Helmholtz free energy of the $v_0$-interaction system in terms of the conventional DFT [6-12].
On the other hand, the functional integration over fluctuating potential field $\phi$ around $i\psi^*$ yields
\begin{flalign}
&e^{-\Delta\Omega[\rho]}=\int D\phi\,e^{-\Delta\mathcal{A}[\rho,\phi]},
\label{appendix hk2}
\end{flalign}
where
\begin{align}
\Delta\mathcal{A}[\rho,\phi]&=\Omega[v_0,-i\phi+\psi^*]-\Omega_0[v_0,\psi^*]+\int d{\bf r}\,i\rho({\bf r})\phi({\bf r}).
\label{appendix additional omega}
\end{align}
In the Gaussian approximation, the additional functional $\Delta\mathcal{A}$ around the saddle-point path $i\psi^*$ is written as
\begin{flalign}
\Delta\mathcal{A}[\rho,\phi]
&=\frac{1}{2}\iint d{\bf r}d{\bf r}'\,
\left.
\phi({\bf r})\frac{\delta^2\Omega_0}{\delta \psi({\bf r})\delta \psi({\bf r}')}
\right|_{\psi=i\psi^*} \phi({\bf r}')\nonumber\\
&=\frac{1}{2}\iint d{\bf r}d{\bf r}'\,\phi({\bf r})\,
G_0({\bf r}-{\bf r}';\rho({\bf r}))\,\phi({\bf r}'),\nonumber\\
G_0({\bf r}-{\bf r}';\rho({\bf r}))&=
\rho({\bf r})\rho({\bf r}')h_0({\bf r}-{\bf r}')+\rho({\bf r})\delta({\bf r}-{\bf r}').
\label{appendix gaussian h phi}
\end{flalign}
It follows from eqs. (\ref{appendix hk2}) and (\ref{appendix gaussian h phi}) that
\begin{equation}
\Delta\Omega=\frac{1}{2}\ln\left|
\mathrm{det}\,G_0({\bf r}-{\bf r}';\rho({\bf r}))\right|.
\label{appendix delta f}
\end{equation}
We thus obtain eqs. (\ref{dfi start}) to (\ref{u1}) from combining eqs. (\ref{appendix conditional free}), (\ref{appendix legendre}) and (\ref{appendix delta f}).

\subsection{Quadratic expansion of $\Omega^*[\rho]$ around the reference density $\rho_{\mathrm{ref}}({\bf r})$: derivation of eqs. (\ref{delta f}), (\ref{g0 h}), and (\ref{expansion4integral})}
The functional Taylor expansion leads to
\begin{flalign}
&\mathcal{F}_0[\rho_{\mathrm{ref}}+n]+U_1[\rho_{\mathrm{ref}}+n]\nonumber\\
&=\mathcal{F}_0[\rho_{\mathrm{ref}}]+U_1[\rho_{\mathrm{ref}}]
+\int d{\bf r}\left.\left\{
\frac{\delta \mathcal{F}_0^{\mathrm{ex}}[\rho]}{\delta\rho({\bf r})}
\right|_{\rho=\rho_{\mathrm{ref}}}
+\left.\frac{\delta \Omega_{\mathrm{att}}^*[\rho]}{\delta\rho({\bf r})}
\right|_{\rho=\rho_{\mathrm{ref}}}
\right\}n({\bf r})\nonumber\\
&\qquad\quad
+\frac{1}{2}\iint d{\bf r}d{\bf r}'\left\{
\left.
\frac{\delta^2 \mathcal{F}_0[\rho]}{\delta\rho({\bf r})\delta\rho({\bf r}')}
\right|_{\rho=\rho_{\mathrm{ref}}}
+\left.
\frac{\delta^2 U_1[\rho]}{\delta\rho({\bf r})\delta\rho({\bf r}')}
\right|_{\rho=\rho_{\mathrm{ref}}}
\right\}
n({\bf r})n({\bf r}')\nonumber\\
&\approx
\mathcal{F}_0[\rho_{\mathrm{ref}}]+U_1[\rho_{\mathrm{ref}}]
+\int d{\bf r}\left.
\frac{\delta \mathcal{F}_0^{\mathrm{ex}}[\rho]}{\delta\rho({\bf r})}
\right|_{\rho=\rho_{\mathrm{ref}}}n({\bf r})
+\frac{1}{2}\iint d{\bf r}d{\bf r}'\left.
\frac{\delta^2 \mathcal{F}_0[\rho]}{\delta\rho({\bf r})\delta\rho({\bf r}')}
\right|_{\rho=\rho_{\mathrm{ref}}}n({\bf r})n({\bf r}').
\label{appendix functional expansion}
\end{flalign}
When reduced to the last line of the above equation (\ref{appendix functional expansion}), we have used the saddle-point equation (\ref{sp attractive}), in addition to eqs. (\ref{conditional}) to (\ref{attractive conditional}), or the equality, $\mathcal{F}_0[\rho]+U_1[\rho]-\int d{\bf r}\rho({\bf r})\mu=\mathcal{F}_0^{\mathrm{ex}}[\rho]+\Omega_{\mathrm{att}}[\rho]$, and have also neglected the quadratic term in the $n$-expansion of $U_1[\rho_{\mathrm{ref}}+n]$ due to the mean-field approximation of attractive interaction energy.
Plugging the RY functional (\ref{our ry}) into eq. (\ref{appendix functional expansion}), eq. (\ref{expansion4integral}) is verified with the approximation that
\begin{flalign}
\left.
\frac{\delta^2 \mathcal{F}_0[\rho]}{\delta\rho({\bf r})\delta\rho({\bf r}')}
\right|_{\rho=\rho_{\mathrm{ref}}}
=G_0^{-1}({\bf r}-{\bf r}';\rho_{\mathrm{ref}}({\bf r}))
&=-c_0({\bf r}-{\bf r}')+\frac{\delta({\bf r}-{\bf r}')}{\rho_{\mathrm{ref}}({\bf r})}\nonumber\\
&\approx
-c_0({\bf r}-{\bf r}')+\frac{\delta({\bf r}-{\bf r}')}{\rho_{\mathrm{ref}}^0}\nonumber\\
&\approx G_0^{-1}({\bf r}-{\bf r}';\rho_{\mathrm{ref}}^0),
\label{appendix g0 approx}
\end{flalign}
where the approximation used in the last line is the same as the approximation in eq. (\ref{underline density}).
By applying the same approximation as eq. (\ref{appendix g0 approx}) to eq. (\ref{appendix delta f}), the approximate logarithmic functional given by eqs. (\ref{delta f}) and (\ref{g0 h}) can also be validated.

\subsection{Transforming eq. (\ref{delta u0}) to eq. (\ref{shifted delta u0}) due to the shift of fluctuating density field}
We express $\Delta \mathcal{F}_0[n]$ given by eq. (\ref{delta u0}) as
\begin{flalign}
&\Delta \mathcal{F}_0[n]
=\frac{1}{2}\iint d{\bf r}d{\bf r}'\,
n({\bf r})\,G_0^{-1}({\bf r}-{\bf r}';\rho_{\mathrm{ref}}^0)n({\bf r}')
-\iint d{\bf r}d{\bf r}'
\,n({\bf r})c_0({\bf r}-{\bf r}')\Delta\rho_{\mathrm{ref}}({\bf r}'),
\nonumber\\
&=\frac{1}{2}\int d{\bf 1}\,n({\bf 1})\left[
\int d{\bf 2}\,G_0^{-1}({\bf 1}-{\bf 2})
\underline{\left\{n({\bf 2})
-2\iint d{\bf 3}\,d{\bf 4}\>G_0({\bf 2}-{\bf 3})c_0({\bf 3}-{\bf 4})\Delta\rho_{\mathrm{ref}}({\bf 4})
\right\}}\right],
\label{appendix delta u0 first}
\end{flalign}
where $G_0({\bf r}-{\bf r}';\rho_{\mathrm{ref}}^0)$ has been abbreviated as $G_0({\bf r}-{\bf r}')$ in the last equality.
The above underlined terms in the last line of eq. (\ref{appendix delta u0 first}) implies the necessity to shift the fluctuating density field from $n({\bf r}) $to $\widetilde{n}({\bf r})$ defined by eq. (\ref{shift n}) for performing the Gaussian integration over the fluctuating density field.

To see this, we rewrite the first term on the rhs of eq. (\ref{shifted delta u0}) using the $n$--field as follows:
\begin{flalign}
&\frac{1}{2}\iint d{\bf 1}d{\bf 2}\,\widetilde{n}({\bf 1})G_0^{-1}({\bf 1}-{\bf 2})\widetilde{n}({\bf 2})\nonumber\\
&=\frac{1}{2}\iint d{\bf 1}d{\bf 2}\,n({\bf 1})G_0^{-1}({\bf 1}-{\bf 2})n({\bf 2})\nonumber\\
&-\frac{1}{2}\iiiint d{\bf 1}\,d{\bf 2}\,d{\bf 3}\,d{\bf 4}\>G_0^{-1}({\bf 1}-{\bf 2})\nonumber\\
&\qquad\times\left\{
n({\bf 1})G_0({\bf 2}-{\bf 3})c_0({\bf 3}-{\bf 4})\Delta\rho_{\mathrm{ref}}({\bf 4})
+n({\bf 2})G_0({\bf 1}-{\bf 3})c_0({\bf 3}-{\bf 4})\Delta\rho_{\mathrm{ref}}({\bf 4})
\right\}\nonumber\\
&+\frac{1}{2}\iiint d{\bf 1}\,d{\bf 2}\,d{\bf 3}\>G_0^{-1}({\bf 1}-{\bf 2})\nonumber\\
&\qquad\times\left[\int d{\bf 4}\,
G_0({\bf 2}-{\bf 3})c_0({\bf 3}-{\bf 4})\Delta\rho_{\mathrm{ref}}({\bf 4})
\right]
\left[\int d{\bf 5}\,
G_0({\bf 1}-{\bf 3})c_0({\bf 3}-{\bf 5})\Delta\rho_{\mathrm{ref}}({\bf 5})
\right]\nonumber\\
&=\frac{1}{2}\iint d{\bf r}d{\bf r}'n({\bf r})G_0^{-1}({\bf r}-{\bf r}')n({\bf r}')
-\iint d{\bf r}d{\bf r}'n({\bf r})c_0({\bf r}-{\bf r}')\Delta\rho_{\mathrm{ref}}({\bf r}')\nonumber\\
&\qquad+\frac{1}{2}\iiiint d{\bf 1}\,d{\bf 3}\,d{\bf 4}\,d{\bf 5}
\Delta\rho_{\mathrm{ref}}({\bf 4})c_0({\bf 4}-{\bf 1})G_0({\bf 1}-{\bf 3})c_0({\bf 3}-{\bf 5})\Delta\rho_{\mathrm{ref}}({\bf 5})\nonumber\\
&=\Delta \mathcal{F}_0[n]
+\frac{1}{2}\iint d{\bf r}d{\bf r}'\Delta\rho_{\mathrm{ref}}({\bf r})\left\{
h_0({\bf r}-{\bf r}')-c_0({\bf r}-{\bf r}')
\right\}\Delta\rho_{\mathrm{ref}}({\bf r}').
\label{appendix delta n quadratic}
\end{flalign}
In obtaining the above last line of eq. (\ref{appendix delta n quadratic}), use has been made of the following derivation due to the Ornstein-Zernike equation (\ref{inhomo oz}):
\begin{flalign}
\iint d{\bf 1}\,d{\bf 3}\,c_0({\bf 4}-{\bf 1})G_0({\bf 1}-{\bf 3})c({\bf 3}-{\bf 5})
&=\iint d{\bf 1}\,d{\bf 3}\,c_0({\bf 4}-{\bf 1})\left\{
\delta({\bf 1}-{\bf 3})+\rho_{\mathrm{ref}}^0h_0({\bf 1}-{\bf 3})\right\}\rho_{\mathrm{ref}}^0
c_0({\bf 3}-{\bf 5})\nonumber\\
&=\int d{\bf 3}\,h_0({\bf 4}-{\bf 3})\rho_{\mathrm{ref}}^0c_0({\bf 3}-{\bf 5})\nonumber\\
&=h_0({\bf 4}-{\bf 5})-c_0({\bf 4}-{\bf 5}).
\label{appendix oz transform}
\end{flalign}
Combination of eqs. (\ref{appendix delta u0 first}) and (\ref{appendix delta n quadratic}) proves eq. (\ref{shifted delta u0}).

\end{document}